\documentclass[12pt]{article}%
\usepackage[nosort]{cite}
\usepackage{graphicx}
\usepackage{multicol}
\usepackage{amsfonts}
\usepackage{amssymb}
\usepackage{amsmath}
\usepackage{heck}
\usepackage{afterpage}
\usepackage{setspace}
\usepackage{verbatim}
\usepackage{color}
\usepackage{longtable}
\usepackage{float}
\usepackage{subcaption}
\usepackage{epsfig}
\usepackage{epstopdf}
\usepackage{adjustbox}
\usepackage{tikz}
\usepackage[margin=1in]{geometry}
\usepackage{titletoc}
\usepackage{hyperref}%
\usepackage{mathrsfs}

\providecommand{\U}[1]{\protect\rule{.1in}{.1in}}
\newsavebox{\mysavebox}

\hypersetup{colorlinks,citecolor=black,filecolor=black,linkcolor=black,urlcolor=black,pdftex}
\usetikzlibrary{decorations.markings}

\numberwithin{equation}{section}

\newcommand{\be}{\begin{equation}}
\newcommand{\ee}{\end{equation}}

\usepackage{tikz}
\usetikzlibrary{arrows}
\usetikzlibrary{arrows.meta}
\usetikzlibrary{arrows,decorations.pathmorphing}
\usetikzlibrary{shapes.geometric,calc,arrows, positioning,shapes.misc,decorations.markings}
\tikzset{
  big arrow/.style={
    decoration={markings,mark=at position 1 with {\arrow[scale=2,#1]{>}}},
    postaction={decorate},
    shorten >=0.4pt},
  big arrow/.default=black}

\pgfdeclarelayer{edgelayer}
\pgfdeclarelayer{nodelayer}
\pgfsetlayers{edgelayer,nodelayer,main}
\tikzstyle{none}=[inner sep=0pt]

\tikzstyle{NodeCross}=[draw, shape=circle, cross out, inner sep=0pt, minimum size=6pt,line width=0.25mm]
\tikzstyle{Circle}=[draw, shape=circle, black, fill=black, inner sep=0pt, minimum size=6pt]
\tikzstyle{circle}=[draw, shape=circle, black, fill=black, inner sep=0pt, minimum size=16pt]
\tikzstyle{Star}=[draw, shape=star, fill=red, star points=8, inner sep=0pt, minimum size=8pt]
\tikzstyle{CircleRed}=[draw, shape=circle, black, fill=red, inner sep=0pt, minimum size=6pt]
\tikzstyle{StarP}=[draw={rgb,255: red,128; green,0; blue,128}, shape=star, fill={rgb,256: red,128; green,0; blue,128}, star points=8, inner sep=0pt, minimum size=12pt]
\tikzstyle{ShadedCircRed}=[draw=red, shape=circle, fill={rgb, 255: red,255; green,114; blue, 118}, inner sep=0pt, minimum size=80pt, line width=0.5mm, fill opacity=0.2]
\tikzstyle{ShadedCircRed2}=[draw=red, shape=circle, fill={rgb, 255: red,255; green,114; blue, 118}, inner sep=0pt, minimum size=10pt]
\tikzstyle{ShadedCircRed3}=[draw=black, shape=rectangle, fill={rgb, 255: red,255; green,114; blue, 118}, inner sep=0pt, minimum size=113pt, line width=0.25mm]
\tikzstyle{ShadedCirc}=[draw=red, shape=circle, fill=white, inner sep=0pt, minimum size=45pt,  fill opacity=1.0,  line width=0.5mm]
\tikzstyle{CircleBlue}=[draw, shape=circle, fill=blue, inner sep=0pt, minimum size=6pt]
\tikzstyle{BigCirclePurple}=[draw, shape=circle, fill={rgb,255: red,191; green,0; blue,191}, inner sep=0pt, minimum size=12pt]
\tikzstyle{CirclePurple}=[draw, shape=circle, fill={rgb,255: red,191; green,0; blue,191}, inner sep=0pt, minimum size=8pt]
\tikzstyle{EmptyCircle}=[draw, shape=circle, inner sep=0pt, minimum size=4pt]
\tikzstyle{GreenCircle}=[draw, shape=circle,  fill={rgb,255: red,80; green,200; blue,120}, inner sep=0pt, minimum size=8pt]
\tikzstyle{BrownCircle}=[draw, shape=circle,  fill={rgb,255: red,210; green,105; blue,30}, inner sep=0pt, minimum size=8pt]
\tikzstyle{CirclePurpleSmall}=[draw, shape=circle, fill={rgb,255: red,191; green,0; blue,191}, inner sep=0pt, minimum size=4pt]
\tikzstyle{BigCircleGreen}=[draw, shape=circle, fill={rgb,255: red,0; green,191; blue,0}, inner sep=0pt, minimum size=12pt]
\tikzstyle{BigCircleBlue}=[draw, shape=circle, fill={rgb,255: red,0; green,0; blue,191}, inner sep=0pt, minimum size=12pt]
\tikzstyle{BigCircleRed}=[draw, shape=circle, fill={rgb,255: red,191; green,0; blue,0}, inner sep=0pt, minimum size=12pt]
\tikzstyle{BrownCircleSmall}=[draw, shape=circle,  fill={rgb,255: red,210; green,105; blue,30}, inner sep=0pt, minimum size=6pt]

\tikzstyle{DashedLine}=[-, densely dashed, line width=0.25mm]
\tikzstyle{DottedLine}=[-, dotted, line width=0.25mm]
\tikzstyle{ThickLine}=[-, line width=0.25mm]
\tikzstyle{ArrowLineRight}=[-, -{Stealth[scale=1.25]}, line width=0.25mm, scale=5]
\tikzstyle{ArrowLineRed}=[-, draw={rgb,255: red,191; green,0; blue,0}, -{Stealth[scale=1.75]}, line width=0.1mm, scale=5]
\tikzstyle{RedLine}=[-, draw={rgb,255: red,191; green,0; blue,0}, fill=none, line width=0.5mm]
\tikzstyle{DashedLineThin}=[-, densely dashed, line width=0.125mm, fill=none, draw=black]
\tikzstyle{DottedRed}=[-, dotted, draw={rgb,255: red,191; green,0; blue,0}, fill=none, line width=0.25mm]
\tikzstyle{DashedRed}=[-, densely dashed, draw={rgb,255: red,191; green,0; blue,0}, fill=none, line width=0.25mm]
\tikzstyle{BlueLine}=[-, draw={rgb,255: red,0; green,0; blue,191}, fill=none, line width=0.5mm]
\tikzstyle{ArrowLineBlue}=[-, draw={rgb,255: red,0; green,0; blue,191}, -{Stealth[scale=1.75]}, line width=0.1mm, scale=5]
\tikzstyle{GreenDoubleArrow}=[<->, draw={rgb,155: red,0; green,255; blue,0},  line width= 0.5mm, scale=5]
\tikzstyle{RedDoubleArrow}=[<->, draw={rgb,255: red,255; green,0; blue,0},  line width= 0.5mm, scale=5]
\tikzstyle{BlueDottedLight}=[-, dotted, draw={rgb,255: red,0; green,0; blue,191}, fill=none, line width=0.3mm]
\tikzstyle{BrownLine}=[-, draw={rgb,255: red,210; green,105; blue,30}, fill=none, line width=0.5mm]
\tikzstyle{DottedRed}=[-, dotted, draw={rgb,255: red,191; green,0; blue,0}, fill=none, dotted, line width=0.5mm]
\tikzstyle{DottedPurple}=[-, dotted, draw={rgb,255: red,191; green,0; blue,191}, fill=none, dotted, line width=0.5mm]
\tikzstyle{BlueDottedLight}=[-, dotted, draw={rgb,255: red,0; green,0; blue,191}, fill=none, line width=0.5mm]
\tikzstyle{ArrowLinePurple}=[-, draw={rgb,255: red,191; green,0; blue,191}, -{Stealth[scale=1.75]}, line width=0.5mm, scale=5]
\tikzstyle{DashedLineGreen}=[-, densely dashed, draw={rgb,255: red,74; green,103; blue,65}, line width=0.25mm]
\tikzstyle{LineGreen}=[-, draw={rgb,255: red, 74; green,200; blue,65}, line width=0.5mm]
\tikzstyle{ArrowLineGreen}=[-, draw={rgb,255: red,0; green,191; blue,0}, -{Stealth[scale=1.75]}, line width=0.5mm, scale=5]
\tikzstyle{GreenLine}=[-, draw={rgb,255: red,0; green,191; blue,0}, fill=none, line width=0.5mm]
\tikzstyle{PurpleLine}=[-, draw={rgb,255: red,191; green,0; blue,191}, fill=none, line width=0.5mm]
\tikzstyle{PPurpleLine}=[-, draw={rgb,255: red,191; green,0; blue,191}, fill=none, line width=2.5mm]
\tikzstyle{DPurpleLine}=[-, dotted, draw={rgb,255: red,191; green,0; blue,191}, fill=none, line width=0.5mm]
\tikzstyle{SBrownLine}=[-, draw={rgb,255: red,191; green,0; blue,191}, fill=none, opacity=0.35, line width=2.5mm]

\tikzset{snake it/.style={decorate, decoration=snake}}

\usetikzlibrary{decorations.markings}

\usetikzlibrary{hobby}

\tikzset{
dashstar/.style={
 dash pattern=on 5pt off 5pt,
 postaction={
  decorate,
  decoration={
   markings,
   mark=between positions 9pt and 1 step 10pt with {
     \node[color=red] {*};
   }
  }
 }
},
dashstarstar/.style={ 
 dash pattern=on 5pt off 10pt,
 postaction={
   decorate,
   decoration={
     markings,
     mark=between positions 10pt and 1
          step 15pt
           with {
            \node at (-2pt,0pt) {\pgfuseplotmark{star}};
            \node at (2pt,0pt) {\pgfuseplotmark{star}};
           }
   }
 }
}
}
\usepackage{pgfplots}
\pgfplotsset{compat=1.16}

\newcommand{\ba}{\begin{aligned}}
\newcommand{\ea}{\end{aligned}}

\begin{document}

\date{January 2024}

\title{On the Holographic Dual of a \\[4mm] Topological Symmetry Operator}

\institution{PENN}{\centerline{$^{1}$Department of Physics and Astronomy, University of Pennsylvania, Philadelphia, PA 19104, USA}}
\institution{PENNmath}{\centerline{$^{2}$Department of Mathematics, University of Pennsylvania, Philadelphia, PA 19104, USA}}

\authors{
Jonathan J. Heckman\worksat{\PENN,\PENNmath}\footnote{e-mail: \texttt{jheckman@sas.upenn.edu}},
Max H\"ubner\worksat{\PENN}\footnote{e-mail: \texttt{hmax@sas.upenn.edu}}, and
Chitraang Murdia\worksat{\PENN}\footnote{e-mail: \texttt{murdia@sas.upenn.edu}}
}

\abstract{We study the holographic dual of a topological symmetry operator in the context of the AdS/CFT correspondence.
Symmetry operators arise from topological field theories localized on a subspace of the boundary CFT spacetime.
We use bottom up considerations to construct the topological sector associated with their bulk counterparts. In particular, by exploiting the
structure of entanglement wedge reconstruction we argue that the bulk counterpart has a non-topological worldvolume action, i.e.,
it describes a dynamical object. As a consequence, we find that there are no global $p$-form symmetries for $p \geq 0$ in asymptotically AdS spacetimes, which includes the case of non-invertible symmetries. Provided one has a suitable notion of subregion-subregion duality, our argument for the absence of bulk global symmetries applies to more general spacetimes.
These considerations also motivate us to consider for general QFTs (holographic or not) the notion of lower-form symmetries, namely, $(-m)$-form symmetries for $m \geq 2$. }

\maketitle

\enlargethispage{\baselineskip}

\setcounter{tocdepth}{2}

\section{Introduction}

Symmetries provide important constraints on observables in
physical systems. An important recent lesson is that global symmetries also
encode a rich topological structure in a quantum field theory (QFT)
\cite{Gaiotto:2014kfa}. In a $D$-dimensional QFT a
topological symmetry operator supported on a closed manifold
$Y$ of dimension $(D-p-1)$ (codimension $p+1$) implements a $p$-form symmetry action on
objects of dimension at least $p$.\footnote{One might prefer to restrict to
ordinary linking, but it is helpful to work more broadly. For example, in a 3D Chern-Simons theory with
charge conjugation symmetry, we have a $0$-form symmetry operator of codimension $2$ which acts on the line operators.}
A priori, the theory on $Y$ could support a non-trivial topological field theory (TFT).
In this broader context, a non-group like fusion rule reflects the appearance of
a non-invertible symmetry.\footnote{See, e.g., the reviews \cite{Cordova:2022ruw, Schafer-Nameki:2023jdn, Bhardwaj:2023kri, Luo:2023ive, Brennan:2023mmt, Shao:2023gho} and references therein.}

Now, in the context of the AdS/CFT correspondence \cite{Maldacena:1997re,
Witten:1998qj, Gubser:1998bc} relating a gravitational theory on
$(D+1)$-dimensional Anti de Sitter (AdS) space and a $D$-dimensional
conformal field theory (CFT), one expects that any operator of the boundary
CFT\ should have a bulk counterpart. Recent progress in the context of stringy
holographic and geometric engineering setups indicates that these topological
operators \textquotedblleft come to life\textquotedblright\ in the bulk as the
topological sector of dynamical branes \cite{Apruzzi:2022rei,
GarciaEtxebarria:2022vzq, Heckman:2022muc, Heckman:2022xgu, Dierigl:2023jdp,
Cvetic:2023plv, Bah:2023ymy, Apruzzi:2023uma, Cvetic:2023pgm}.

Our aim in this note will be to give a bottom up explanation for some of these
observations without reference to a specific top down construction.

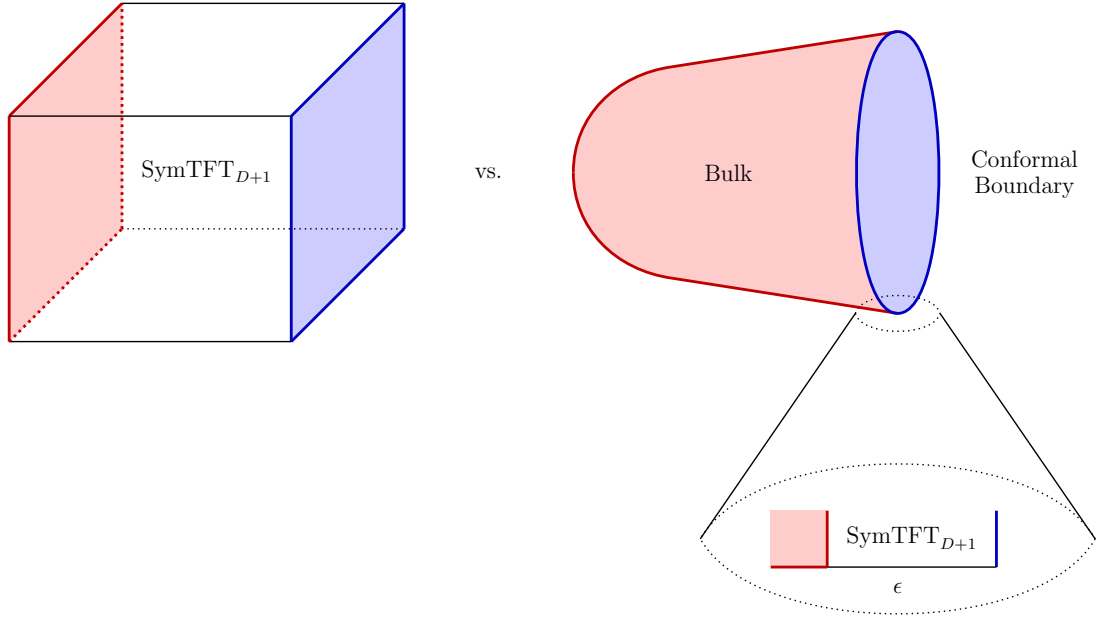
\begin{figure}[h]
\centering
\scalebox{0.75}{
\begin{tikzpicture}
	\begin{pgfonlayer}{nodelayer}
		\node [style=none] (0) at (-9, -2) {};
		\node [style=none] (1) at (-9, 2) {};
		\node [style=none] (2) at (-7, 4) {};
		\node [style=none] (3) at (-7, 0) {};
		\node [style=none] (4) at (-4, -2) {};
		\node [style=none] (5) at (-2, 0) {};
		\node [style=none] (6) at (-2, 4) {};
		\node [style=none] (7) at (-4, 2) {};
		\node [style=none] (8) at (-5.5, 1) {$\mathrm{SymTFT}_{D+1}$};
		\node [style=none] (9) at (-8.5, 3.5) {};
		\node [style=none] (10) at (-2.5, -1.5) {};
		\node [style=none] (11) at (6.75, 3.5) {};
		\node [style=none] (12) at (6.75, -1.5) {};
		\node [style=none] (13) at (1, 1) {};
		\node [style=none] (14) at (3.75, 1) {Bulk};
		\node [style=none] (15) at (9, 1.25) {Conformal};
		\node [style=none] (16) at (-0.5, 1) {vs.};
		\node [style=none] (17) at (7.5, -1.5) {};
		\node [style=none] (18) at (7.5, -1.5) {};
		\node [style=none] (19) at (10.25, -5.5) {};
		\node [style=none] (20) at (6, -1.5) {};
		\node [style=none] (23) at (3.25, -5.5) {};
		\node [style=none] (24) at (8.5, -5) {};
				\node [style=none] (26) at (8.5, -6) {};
		\node [style=none] (27) at (5.5, -5) {};
				\node [style=none] (29) at (5.5, -6) {};
		\node [style=none] (30) at (6.75, -6.375) {$\epsilon$};
		\node [style=none] (31) at (7, -5.5) {$\mathrm{SymTFT}_{D+1}$};
		\node [style=none] (32) at (4.5, -6) {};
		\node [style=none] (34) at (9, 0.75) {Boundary};
		\node [style=none] (35) at (-11, -5) {};
		\node [style=none] (37) at (2.75, 2.875) {};
		\node [style=none] (38) at (2.75, -0.875) {};
		\node [style=none] (42) at (10.5, -1) {};
	\end{pgfonlayer}
	\begin{pgfonlayer}{edgelayer}
		\filldraw[fill=red!20, draw=red!20]  (2.75, 2.875) -- (2.75, -0.875) -- (6.75, -1.5) -- (6.75, 3.5) -- cycle;
		\filldraw[fill=red!20, draw=red!20]  (-9, -2) -- (-9, 2) -- (-7, 4) -- (-7, 0) -- cycle;
		\filldraw[fill=red!20, draw=red!20]  (5.5, -6) -- (4.5, -6) -- (4.5, -5) -- (5.5, -5) -- cycle;
				\filldraw[fill=blue!20, draw=blue!20]  (-4, -2) -- (-4, 2) -- (-2, 4) -- (-2, 0) -- cycle;
		\filldraw[fill=blue!20, draw=blue!20] (6.75, 1) ellipse (0.75cm and 2.5cm);
		\filldraw[fill=red!20, draw=red!20] (2.85, 1) ellipse (1.82cm and 1.85cm);
		\draw [style=RedLine] (0.center) to (1.center);
		\draw [style=RedLine] (1.center) to (2.center);
		\draw [style=BlueLine] (7.center) to (6.center);
		\draw [style=BlueLine] (7.center) to (4.center);
		\draw [style=ThickLine] (0.center) to (4.center);
		\draw [style=ThickLine] (1.center) to (7.center);
		\draw [style=ThickLine] (2.center) to (6.center);
		\draw [style=DottedLine] (3.center) to (5.center);
		\draw [style=DottedRed] (0.center) to (3.center);
		\draw [style=DottedRed] (3.center) to (2.center);
		\draw [style=BlueLine] (4.center) to (5.center);
		\draw [style=BlueLine] (5.center) to (6.center);
						\draw [style=ThickLine] (17.center) to (19.center);
		\draw [style=ThickLine] (23.center) to (20.center);
		\draw [style=DottedLine, bend left=75, looseness=0.75] (20.center) to (18.center);
		\draw [style=DottedLine, bend right=75, looseness=0.75] (20.center) to (18.center);
		\draw [style=DottedLine, bend right=60, looseness=0.75] (19.center) to (23.center);
		\draw [style=DottedLine, bend left=60, looseness=0.75] (19.center) to (23.center);
		\draw [style=BlueLine] (24.center) to (26.center);
				\draw [style=RedLine] (27.center) to (29.center);
				\draw [style=ThickLine] (26.center) to (29.center);
		\draw [style=RedLine] (29.center) to (32.center);
		\draw [style=RedLine] (12.center) to (38.center);
		\draw [style=RedLine, in=-90, out=172] (38.center) to (13.center);
		\draw [style=RedLine, in=90, out=-172] (37.center) to (13.center);
		\draw [style=RedLine] (37.center) to (11.center);
		\draw [style=BlueLine, bend right=90, looseness=0.50] (11.center) to (12.center);
		\draw [style=BlueLine, bend left=90, looseness=0.50] (11.center) to (12.center);
	\end{pgfonlayer}
\end{tikzpicture}}\caption{LHS: We depict the standard SymTFT sandwich. The
SymTFT$_{D+1}$ is supported on a slab of dimension $D+1$ with
boundaries of dimension $D$ at which physical and topological boundary
conditions are imposed as indicated by the shaded regions. RHS: In holographic
setups the full bulk is dual to the physical theory and therefore sets the
physical boundary conditions, topological boundary conditions are imposed
asymptotically at the conformal boundary. The SymTFT$_{D+1}$ can be thought of as
supported between these on a slab / sliver of width $\epsilon\rightarrow0$. The physical
boundary is now determined by a gravitational system in dimension $D+1$.}%
\label{fig:SymTFTvsADSCFT}%
\end{figure}

A helpful clue for how to proceed is the associated symmetry topological field
theory (SymTFT) which governs the global symmetries of a QFT.\footnote{See, e.g., references
\cite{Reshetikhin:1991tc, Turaev:1992hq, Barrett:1993ab, Witten:1998wy,
Fuchs:2002cm, Kirillov:2010nh, Kapustin:2010if, Kitaev:2011dxc, Fuchs:2012dt,
Freed:2012bs, Freed:2018cec, Apruzzi:2021nmk, Freed:2022qnc, Kaidi:2022cpf,
Baume:2023kkf,Brennan:2024fgj}.} For a $D$-dimensional QFT with a given set of categorical
symmetries, there is a $(D+1)$-dimensional SymTFT$_{D+1}$ which
governs the global form of the QFT$_D$; this involves extending the $D$-dimensional
QFT by an interval; at one boundary we have the
(relative) QFT, and at the other end we introduce topological /
gapped boundary conditions to specify the global form of the QFT.\footnote{The structure of the SymTFT is best understood
in the case of finite and discrete symmetries. In the case of continuous symmetries, there are some subtleties with
demanding a purely gapped bulk, and to a certain extent it is not necessary (and sometimes undesirable)
to enforce this structure since the ``global form'' for continuous symmetries is less ambiguous; we simply need to enforce suitable
boundary conditions for continuous gauge fields near a boundary. For a recent discussion on SymTFTs for continuous symmetries see reference \cite{Brennan:2024fgj}. Let us also comment that in all of these circumstances we can still consider a bulk TFT which detects the anomalies of a boundary theory which by (a mild) abuse of terminology we shall refer to as the SymTFT.}
This is of course highly reminiscent of the AdS/CFT\ correspondence, although in the
latter, we really only have the single conformal boundary of AdS; to some extent
the physical boundary conditions of the
SymTFT have instead been \textquotedblleft smeared\textquotedblright\ over
the entire bulk (as it must, since the gravitational theory \textquotedblleft
knows\textquotedblright\ about all the local interactions of the CFT!). See
figure \ref{fig:SymTFTvsADSCFT} for a depiction of the two bulk / boundary correspondences.

Another clue is the recent argument presented in \cite{Harlow:2018jwu,
Harlow:2018tng} that certain global symmetries cannot be present in
asymptotically AdS spacetimes.\footnote{The general expectation is that in a theory of quantum gravity there are no global symmetries. See e.g., references \cite{Misner:1957mt, Banks:1988yz, Susskind:1995da, Polchinski:2003bq, Banks:2010zn}.} The argument follows via \textquotedblleft
proof by contradiction,\textquotedblright\ showing that any topological
operator of the boundary theory cannot reach \textquotedblleft far
enough\textquotedblright\ into the bulk to influence quasi-localized bulk
operators.\footnote{It is important to note that this proof does not directly
construct a candidate dual for a topological symmetry operator of the boundary
theory. Additionally, there are some technical complications in extending this
discussion to the case of non-invertible symmetries, where the fusion rules
for symmetry operators fail to obey a group-like product rule.} Indeed, the
essential point in this line of reasoning is that entanglement wedges have
only finite extent in the bulk (see e.g., \cite{Ryu:2006bv, Ryu:2006ef,
Hubeny:2007xt}). A final complication is that the arguments presented in
\cite{Harlow:2018jwu,Harlow:2018tng} appear to require $D \geq p + 2$.

While intuitively appealing, this line of reasoning also raises some questions.
For one, the topological nature of the symmetry operators suggests that they can be
deformed arbitrarily, provided linking between operators is maintained. From
this perspective, one reaches a puzzle:\ how could the gravity dual of a
topological operator \textquotedblleft know\textquotedblright\ about the
location of something as metric dependent as an entanglement wedge in the
first place?

Here we address some of these issues, showing that
bottom up considerations concerning bulk reconstruction / holographic RG flows
are enough to argue that bulk symmetry operators are never purely topological,
and so there are no global symmetries. Our line of reasoning shares some
similarities with that presented in \cite{Harlow:2018jwu, Harlow:2018tng} but
is also complementary in many aspects.

With this aim in mind, we consider for $D \geq 2$, a CFT$_D$ with a semi-classical gravity dual,
but we do not assume that global symmetries of the boundary theory are gauged
in the bulk.\footnote{We expect our considerations to also apply for $D = 1$, but in
this case there is not much to discuss other than $(-1)$-form symmetries, which involve parameters of the theory. In known AdS$_2$/CFT$_1$ pairs one often has to entertain an ensemble average over parameters right from the start (see e.g., \cite{Maldacena:2016hyu}) so the analysis will be somewhat different. We defer this case to future work.} Rather, we argue that bulk reconstruction means that symmetry operators must have a bulk dual which sources stress energy in the sense that these operators are now
sensitive to small fluctuations in the metric. As such, they cannot be purely topological. While detailed properties of the resulting object in gravity are model dependent, we can also extract some qualitative properties of the resulting brane action worldvolume theory. Summarizing, the existence of a symmetry operator in the boundary CFT$_D$ allows us to \textit{predict} the existence of a brane in the gravitational bulk.

For completeness, we also revisit the no global symmetries proof given in
\cite{Harlow:2018jwu, Harlow:2018tng}, showing how to also cover the case of
non-invertible symmetries which fuse to condensation defects.
We also comment on the construction of
$(-m)$-form symmetries of a general QFT for $m \geq 2$,
as well as their holographic dual
description in the context of the AdS/CFT\ correspondence.
We conclude by briefly discussing some potential
extensions to more general spacetimes.

\section{Bulk Dual of a Symmetry Operator}

To begin, we consider a CFT\ on a $D$-dimensional spacetime, and assume that
it\ admits a topological symmetry operator $\mathcal{N}$ supported on a closed
$q$-manifold $Y$. This is a codimension $D-q$ symmetry operator and so
specifies a $p$-form symmetry with $p + q = D-1$. For $\mathcal{N}$ to be a
symmetry operator, we require that it commute with the stress tensor
$T_{\mu\nu}$ of the CFT. In general, $\mathcal{N}$ is specified by a
non-trivial topological field theory (TFT) supported on $Y$ so we schematically write this
as:\footnote{We keep implicit the specific symmetry generator, as implemented
by the TFT.}%
\begin{equation}
\mathcal{N}\sim\int[da]\exp\left(  2\pi i\underset{Y}{\int}\mathcal{L}%
_{\text{TFT}}\right)  , \label{eq:OnBoundary}%
\end{equation}
in the obvious notation. It is topological because local variations in the shape of $Y$ do not alter the action of the topological defect on operators of the QFT. One can view this as being enforced by the requirement that the TFT on $Y$ is independent of the spacetime metric for the CFT.

As is the case for any defect, the bulk fields
of the CFT\ specify background fields / sources for the TFT\ action. We say
that an operator $\mathcal{O}$ supported on a subspace of dimension $k\geq p$
of the CFT\ is charged under this symmetry operator when it can non-trivially
intersect / link with the symmetry operator.\footnote{Note, for example, that
a line operator can thus be charged under both a 1-form symmetry as well as a
0-form symmetry.} The action of the symmetry operator on $\mathcal{O}$ will be
denoted as $\mathcal{O}\mapsto\mathcal{O}^{(\mathcal{N)}}$. We first consider
$p\geq0$, returning to $p=-1$ later. Recall that a global $(-1)$-form symmetry is
associated with picking parameters of the CFT.

We now make the further assumption that our CFT\ has a
semi-classical gravity dual.\footnote{A loose form of holography
would assert that any QFT\ should have some gravity dual. In practice this
statement has little content / practical value.} In what follows we shall not
assume that the global symmetry of the boundary theory is gauged. Rather, we shall
ask whether having a global symmetry in the bulk is compatible with bulk reconstruction,
reaching a contradiction.

We would like to characterize the bulk object corresponding to $\mathcal{N}$, as well
as its possible linking with $\mathcal{O}$. Our general expectation is that to
characterize the bulk dual, we ought to convolve a
boundary CFT\ operator with a suitable smearing kernel and its improvement /
reinterpretation in terms of a quantum error correcting code (QECC)
\cite{Hamilton:2006az, Almheiri:2014lwa} (see e.g.,
\cite{DeJonckheere:2017qkk} for a review); objects close to the conformal
boundary will be sharply localized but as we move deeper into the interior of
AdS, these bulk dual objects will become more spread out. For example, in the
case of $\mathcal{O}=\mathcal{O}(x)$ a local operator, we have a smearing
kernel $\mathcal{K}(x^{\prime},x;z)$ so that the resulting bulk operator is of
the form (see e.g., \cite{Hamilton:2006az, Jafferis:2015del,
Pastawski:2015qua, Dong:2016eik, Cotler:2017erl}):
\begin{equation}
\widetilde{\mathcal{O}}(x,z)\sim\int dx^{\prime}\text{ }\mathcal{O}(x^{\prime
})\mathcal{K}(x^{\prime},x;z)
\end{equation}
where $z$ denotes the local radial coordinate in the Poincar\'{e} patch, i.e., we have the empty AdS metric:
\begin{equation}
ds^2 = \ell_{\mathrm{AdS}}^2 \frac{ds^{2}_{\mathrm{CFT}} + dz^2}{z^2},
\end{equation}
with $ds^2_{\mathrm{CFT}}$ the metric of the boundary CFT.
Similar considerations apply for extended operators of the CFT, i.e., they also smear
out in the bulk. We write this as a convolution:
\begin{equation}
\widetilde{\mathcal{O}}\sim\mathcal{O\ast K}\text{.}%
\end{equation}

Let us now turn to the main focus of this note: the bulk dual of a topological symmetry operator $\mathcal{N}$ of the CFT$_D$.
In keeping with our discussion of smearing for CFT operators, we shall refer to the putative bulk dual as $\widetilde{\mathcal{N}}$.
There are two issues one can immediately raise. First of all, how do we know that $\widetilde{\mathcal{N}}$ even exists, and moreover, should we expect it to be topological in the bulk?\footnote{We thank J. McNamara and H. Ooguri for helpful questions and comments on this point.}

First of all, there are good reasons to expect that some object such as $\widetilde{\mathcal{N}}$ does exist in the bulk.
For one, note that for a general QFT$_D$ in $D$ spacetime dimensions we can speak of the associated SymTFT$_{D+1}$. In this picture,
the SymTFT$_{D+1}$ is defined on a slab with one boundary carrying gapped boundary conditions and the other supporting physical boundary conditions, i.e., a relative QFT. Defects of the boundary relative theory are extended by one dimension, and symmetry operators simply ``pull off'' of the boundary, thus maintaining linking in both the boundary and the bulk. So, at least in the SymTFT there is a natural bulk object which we shall refer to as $\mathcal{N}^{\text{stft}}$. An additional comment here is that $\mathcal{N}$ is constructed from fields of the QFT$_D$ so smearing of these fields should in principle produce an object in any putative bulk dual.

Consider next the issue of whether $\widetilde{\mathcal{N}}$ is topological.
To see why this issue is so subtle, suppose we consider integrating a candidate
smearing kernel against some $\mathcal{N}$ supported on a subspace $Y$.
Provided we remain close enough to the boundary so that any non-trivial topological features in the bulk are avoided,
the only possible dual is some object supported on a manifold $Y(z)$ homotopic to $Y$.
On the other hand, precisely because $\mathcal{N}$ is topological,
convolution with a smearing kernel ought to have no effect on $\widetilde{\mathcal{N}}$ at all!

With this motivation in mind, our aim will be to establish not only that $\widetilde{\mathcal{N}}$ exists, but also that it is not topological.
There are two conflicting intuitions which we will need to reconcile: on the one hand
``smearing'' with a topological operator would seem to produce no effect at all. On the other hand, linking between $\mathcal{N}$ and
$\mathcal{O}$ will eventually get smeared out, making it difficult to separately reconstruct $\mathcal{N}$ and $\mathcal{O}$.

Once we establish that $\widetilde{\mathcal{N}}$ exists, we also know that in the $z \rightarrow 0 $ limit it must reduce to $\mathcal{N}$.
Since $\mathcal{N}$ is topological, we can deduce that near the conformal
boundary of AdS, the worldvolume theory on a radial slice
$\widetilde{\mathcal{N}}(z)$ will have to be of the form:\footnote{The TFT here is specified by the boundary theory. There can often be difficulties in constructing this TFT purely in terms of bulk objects; we discuss this issue when we turn to examples with continuous non-abelian symmetries.}
\begin{equation}
\widetilde{\mathcal{N}}(z)\sim \mathcal{Z} \times \int[da]\exp\left(  2\pi i\underset{Y(z)}{\int%
}\mathcal{L}_{\text{TFT}} + \text{Non-Topological}\right)  , \label{Ntildez}%
\end{equation}
where $Y(z)$ is homotopic to $Y(0)=Y$. This is simply because any conventional smearing kernel cannot alter the TFT worldvolume terms.
On the other hand, it could happen that once we move into the bulk there are additional
non-topological terms which depend on local fluctuations
of the bulk metric.\footnote{One
can of course entertain topological couplings such as the Pontryagin density,
but these do not produce a source of stress energy.} We will shortly argue
that the worldvolume action used to specify $\widetilde{\mathcal{N}}(z)$
is non-topological. As such, a global symmetry in the CFT
cannot remain a global symmetry in the bulk. In addition, we have included the possibility of a further dressing term $\mathcal{Z}$ as associated with a theory supported on a chain which stretches from $Y(z)$ to $Y(0)$. This only occurs in situations where the naive SymTFT extrapolation of $\mathcal{L}_{\text{TFT}}$ to the bulk is, on its own, ill-defined.

For each radial slice of the local AdS geometry, we expect that
$\widetilde{\mathcal{O}}(z)$ still links with $\widetilde{\mathcal{N}}(z)$.
From the perspective of the dual CFT, this is expected:\ if we introduce a
small UV\ cutoff length scale, we induce an RG\ flow and so we can track, as a
function of RG\ time, the action of topological symmetry operators on CFT
operators.\footnote{Recall that the practical implementation of holographic
RG requires us to move the CFT a ``small amount'' into the
interior to initiate a flow. See e.g., \cite{deBoer:1999tgo, Bianchi:2001kw}
as well as the review \cite{Skenderis:2002wp}.} The
RG\ flow as we proceed into the infrared could involve non-trivial operator
mixing / transport into the bulk, so there is an \textquotedblleft
RG\ line\textquotedblright\ which extends out from $\widetilde{\mathcal{O}%
}(z)$ back to the boundary at $z=0$. Likewise, we can consider the full
evolution of $\widetilde{\mathcal{N}}(z)$ back to the boundary at $z=0$.
Observe that the linking dimension for $\mathcal{O}$ and $\mathcal{N}$ in the
CFT is different from that of their smeared counterparts
$\widetilde{\mathcal{O}}$ and $\widetilde{\mathcal{N}}$ in the bulk (which is
higher-dimensional). This does not directly imply a contradiction since along
each radial slice we still observe a linking.

In the next two subsections we argue that $\widetilde{\mathcal{N}}(z)$ exists, and moreover, it cannot be purely topological.
To illustrate the main idea we shall go through the argument twice, once in Euclidean signature, where all operator linking / RG statements are on the same footing, and again in Lorentzian signature, where the
physical interpretation of smearing and bulk reconstruction is more apparent.

This will be enough to also establish the absence of bulk global symmetries.
Indeed, given a putative topological symmetry operator $\mathcal{N}^{\mathrm{AdS}}$
for a global symmetry of the bulk, pushing it to the boundary would produce a
topological symmetry operator $\mathcal{N}$ for a global symmetry of the boundary CFT.
Pulling it back into the bulk via bulk reconstruction would then yield
a non-topological $\widetilde{\mathcal{N}} = \mathcal{N}^{\mathrm{AdS}}$. This is
a contradiction, since on the one hand $\mathcal{N}^{\mathrm{AdS}}$ as a global symmetry operator
should not depend on local metric fluctuations, but on the other hand its counterpart $\widetilde{\mathcal{N}}$ does depend on such fluctuations. The contradiction implies the absence of bulk global symmetries.

Finally, let us comment that while the style of our argument is geared towards a ``proof by contradiction''
namely we assume at the outset that the bulk has a global symmetry and then derive a contradiction, the argument carries through
in the same fashion even if we assume that symmetry in the bulk is gauged. This is because we can compute (via the use of a
connection / Wilson line) the difference of $\widetilde{\mathcal{O}}$ and its smeared counterpart. There is still a contribution to the stress energy tensor in this case, and this again establishes that $\widetilde{\mathcal{N}}$ is not topological.

\subsection{Euclidean Signature Analysis}

To begin, suppose our CFT is formulated on a Euclidean signature
manifold of dimension $D$. We can view this as preparing a
specific state of the Lorentzian signature theory. In this case, the empty
AdS geometry is topologically a $(D+1)$-ball; the radial direction of the ball
can be viewed as the RG time of the Euclidean CFT (after introducing a short distance cutoff).

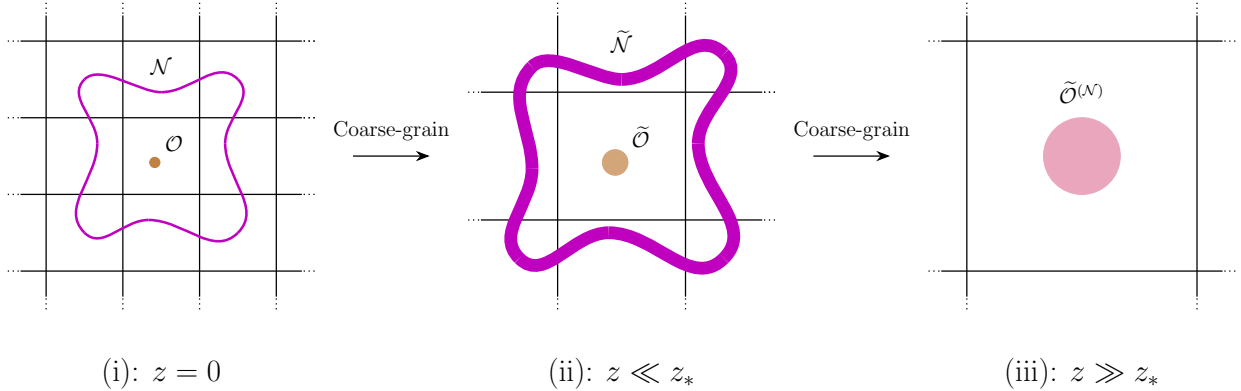
\begin{figure}
\centering
\scalebox{0.68}{
\begin{tikzpicture}
	\begin{pgfonlayer}{nodelayer}
		\node [style=none] (0) at (-2.75, 2.75) {};
		\node [style=none] (1) at (-1.25, 2.75) {};
		\node [style=none] (2) at (0.25, 2.75) {};
		\node [style=none] (3) at (1.75, 2.75) {};
		\node [style=none] (5) at (2.25, 2.25) {};
		\node [style=none] (6) at (2.25, 0.75) {};
		\node [style=none] (7) at (2.25, -0.75) {};
		\node [style=none] (8) at (2.25, -2.25) {};
		\node [style=none] (9) at (1.75, -2.75) {};
		\node [style=none] (10) at (0.25, -2.75) {};
		\node [style=none] (11) at (-1.25, -2.75) {};
		\node [style=none] (12) at (-2.75, -2.75) {};
		\node [style=none] (13) at (-3.25, 2.25) {};
		\node [style=none] (14) at (-3.25, 0.75) {};
		\node [style=none] (15) at (-3.25, -0.75) {};
		\node [style=none] (16) at (-3.25, -2.25) {};
		\node [style=none] (17) at (-2.75, 3) {};
		\node [style=none] (18) at (-1.25, 3) {};
		\node [style=none] (19) at (0.25, 3) {};
		\node [style=none] (20) at (1.75, 3) {};
		\node [style=none] (25) at (1.75, -3) {};
		\node [style=none] (26) at (0.25, -3) {};
		\node [style=none] (27) at (-1.25, -3) {};
		\node [style=none] (28) at (-2.75, -3) {};
		\node [style=none] (29) at (-3.5, -2.25) {};
		\node [style=none] (30) at (-3.5, -0.75) {};
		\node [style=none] (31) at (-3.5, 0.75) {};
		\node [style=none] (32) at (-3.5, 2.25) {};
		\node [style=none] (34) at (-2, 1.5) {};
		\node [style=none] (35) at (-0.5, 1.25) {};
		\node [style=none] (36) at (1, 1.5) {};
		\node [style=none] (37) at (-2, -1.5) {};
		\node [style=none] (38) at (1, -1.5) {};
		\node [style=none] (39) at (0.75, 0.25) {};
		\node [style=none] (40) at (-0.75, -1.25) {};
		\node [style=none] (41) at (-1.75, 0.25) {};
		\node [style=none] (42) at (15.25, 2.75) {};
		\node [style=none] (45) at (19.75, 2.75) {};
		\node [style=none] (46) at (20.25, 2.25) {};
		\node [style=none] (49) at (20.25, -2.25) {};
		\node [style=none] (50) at (19.75, -2.75) {};
		\node [style=none] (53) at (15.25, -2.75) {};
		\node [style=none] (54) at (14.75, 2.25) {};
		\node [style=none] (57) at (14.75, -2.25) {};
		\node [style=none] (58) at (15.25, 3) {};
		\node [style=none] (69) at (15.25, -3) {};
		\node [style=none] (70) at (14.5, -2.25) {};
		\node [style=none] (73) at (14.5, 2.25) {};
		\node [style=none] (75) at (3.25, 0) {};
		\node [style=none] (76) at (4.75, 0) {};
		\node [style=none] (77) at (4, 0.5) {Coarse-grain};
		\node [style=none] (78) at (17.5, 1.25) {$\widetilde{\mathcal{O}}^{(\mathcal{N})}$};
		\node [style=none] (79) at (-0.5, -4.25) {\Large (i): $z=0$};
		\node [style=none] (80) at (8.5, -5) {};
		\node [style=none] (81) at (17.5, -4.25) {\Large (iii): $z\gg z_*$};
		\node [style=none] (82) at (-0.25, 0.25) {$\mathcal{O}$};
		\node [style=none] (83) at (-0.5, 1.75) {$\mathcal{N}$};
		\node [style=none] (84) at (2.5, 2.25) {};
		\node [style=none] (85) at (2.5, 0.75) {};
		\node [style=none] (86) at (2.5, -0.75) {};
		\node [style=none] (87) at (2.5, -2.25) {};
		\node [style=none] (88) at (19.75, 3) {};
		\node [style=none] (89) at (20.5, 2.25) {};
		\node [style=none] (90) at (20.5, -2.25) {};
		\node [style=none] (91) at (19.75, -3) {};
		\node [style=none] (93) at (7.25, 2.75) {};
		\node [style=none] (94) at (9.75, 2.75) {};
		\node [style=none] (97) at (11.25, 1.25) {};
		\node [style=none] (98) at (11.25, -1.25) {};
		\node [style=none] (101) at (9.75, -2.75) {};
		\node [style=none] (102) at (7.25, -2.75) {};
		\node [style=none] (105) at (5.75, 1.25) {};
		\node [style=none] (106) at (5.75, -1.25) {};
		\node [style=none] (109) at (7.25, 3) {};
		\node [style=none] (110) at (9.75, 3) {};
		\node [style=none] (113) at (9.75, -3) {};
		\node [style=none] (114) at (7.25, -3) {};
		\node [style=none] (117) at (5.5, -1.25) {};
		\node [style=none] (118) at (5.5, 1.25) {};
		\node [style=none] (121) at (6.75, 1.75) {};
		\node [style=none] (122) at (8.5, 1.5) {};
		\node [style=none] (123) at (10.5, 2) {};
		\node [style=none] (124) at (6.5, -2) {};
		\node [style=none] (125) at (10.5, -2) {};
		\node [style=none] (126) at (10, 0.25) {};
		\node [style=none] (127) at (8.25, -1.5) {};
		\node [style=none] (128) at (6.75, -0.25) {};
		\node [style=none] (129) at (8.5, -4.25) {\Large (ii): $z\ll z_*$};
		\node [style=none] (130) at (8.875, 0.375) {$\widetilde{\mathcal{O}}$};
		\node [style=none] (131) at (8.5, 2.25) {$\widetilde{\mathcal{N}}$};
		\node [style=none] (133) at (11.5, 1.25) {};
		\node [style=none] (134) at (11.5, -1.25) {};
		\node [style=none] (135) at (12.25, 0) {};
		\node [style=none] (136) at (13.75, 0) {};
		\node [style=none] (137) at (13, 0.5) {Coarse-grain};
	\end{pgfonlayer}
	\begin{pgfonlayer}{edgelayer}
		\filldraw[fill=brown!100, draw=brown!100] (-0.625, -0.125) ellipse (0.1cm and 0.1cm);
		\filldraw[fill=brown!70, draw=brown!70] (8.375, -0.125) ellipse (0.25cm and 0.25cm);
		\filldraw[fill=purple!35, draw=purple!35] (17.5, 0) ellipse (0.75cm and 0.75cm);
		\draw [style=ThickLine] (0.center) to (12.center);
		\draw [style=ThickLine] (1.center) to (11.center);
		\draw [style=ThickLine] (2.center) to (10.center);
		\draw [style=ThickLine] (3.center) to (9.center);
		\draw [style=ThickLine] (8.center) to (16.center);
		\draw [style=ThickLine] (15.center) to (7.center);
		\draw [style=ThickLine] (14.center) to (6.center);
		\draw [style=ThickLine] (5.center) to (13.center);
		\draw [style=DottedLine] (17.center) to (0.center);
		\draw [style=DottedLine] (18.center) to (1.center);
		\draw [style=DottedLine] (19.center) to (2.center);
		\draw [style=DottedLine] (20.center) to (3.center);
		\draw [style=DottedLine] (26.center) to (10.center);
		\draw [style=DottedLine] (27.center) to (11.center);
		\draw [style=DottedLine] (28.center) to (12.center);
		\draw [style=DottedLine] (29.center) to (16.center);
		\draw [style=DottedLine] (30.center) to (15.center);
		\draw [style=DottedLine] (31.center) to (14.center);
		\draw [style=DottedLine] (32.center) to (13.center);
		\draw [style=PurpleLine, in=180, out=45, looseness=0.75] (34.center) to (35.center);
		\draw [style=PurpleLine, in=135, out=0] (35.center) to (36.center);
		\draw [style=PurpleLine, in=90, out=-45] (36.center) to (39.center);
		\draw [style=PurpleLine, in=45, out=-90] (39.center) to (38.center);
		\draw [style=PurpleLine, in=0, out=-135] (38.center) to (40.center);
		\draw [style=PurpleLine, in=-45, out=-180] (40.center) to (37.center);
		\draw [style=PurpleLine, in=-90, out=135] (37.center) to (41.center);
		\draw [style=PurpleLine, in=-135, out=90] (41.center) to (34.center);
		\draw [style=ThickLine] (42.center) to (53.center);
		\draw [style=ThickLine] (45.center) to (50.center);
		\draw [style=ThickLine] (49.center) to (57.center);
		\draw [style=ThickLine] (46.center) to (54.center);
		\draw [style=DottedLine] (58.center) to (42.center);
		\draw [style=DottedLine] (69.center) to (53.center);
		\draw [style=DottedLine] (70.center) to (57.center);
		\draw [style=DottedLine] (73.center) to (54.center);
		\draw [style=ArrowLineRight] (75.center) to (76.center);
		\draw [style=DottedLine] (84.center) to (5.center);
		\draw [style=DottedLine] (85.center) to (6.center);
		\draw [style=DottedLine] (84.center) to (5.center);
		\draw [style=DottedLine] (85.center) to (6.center);
		\draw [style=DottedLine] (86.center) to (7.center);
		\draw [style=DottedLine] (87.center) to (8.center);
		\draw [style=DottedLine] (25.center) to (9.center);
		\draw [style=DottedLine] (88.center) to (45.center);
		\draw [style=DottedLine] (89.center) to (46.center);
		\draw [style=DottedLine] (90.center) to (49.center);
		\draw [style=DottedLine] (91.center) to (50.center);
		\draw [style=ThickLine] (93.center) to (102.center);
		\draw [style=ThickLine] (94.center) to (101.center);
		\draw [style=ThickLine] (106.center) to (98.center);
		\draw [style=ThickLine] (105.center) to (97.center);
		\draw [style=DottedLine] (109.center) to (93.center);
		\draw [style=DottedLine] (110.center) to (94.center);
		\draw [style=DottedLine] (113.center) to (101.center);
		\draw [style=DottedLine] (114.center) to (102.center);
		\draw [style=DottedLine] (117.center) to (106.center);
		\draw [style=DottedLine] (118.center) to (105.center);
		\draw [style=SBrownLine, in=180, out=45, looseness=0.75] (121.center) to (122.center);
		\draw [style=SBrownLine, in=135, out=0] (122.center) to (123.center);
		\draw [style=SBrownLine, in=90, out=-45] (123.center) to (126.center);
		\draw [style=SBrownLine, in=45, out=-90] (126.center) to (125.center);
		\draw [style=SBrownLine, in=0, out=-135] (125.center) to (127.center);
		\draw [style=SBrownLine, in=-45, out=-180] (127.center) to (124.center);
		\draw [style=SBrownLine, in=-90, out=135] (124.center) to (128.center);
		\draw [style=SBrownLine, in=-135, out=90] (128.center) to (121.center);
		\draw [style=DottedLine] (133.center) to (97.center);
		\draw [style=DottedLine] (133.center) to (97.center);
		\draw [style=DottedLine] (134.center) to (98.center);
		\draw [style=ArrowLineRight] (135.center) to (136.center);
	\end{pgfonlayer}
\end{tikzpicture}}
\caption{(i): At zero RG time $z=0$ the operators $\mathcal{N}$ and $\mathcal{O}$ are not smeared. (ii): At small RG times $z\ll z_*$, some smearing has occurred but an observer can still probe the symmetry operator $\widetilde{\mathcal{N}}$ and the operator $\widetilde{\mathcal{O}}$ individually. We denote this as $\widetilde{\mathcal{O}}\oplus \widetilde{\mathcal{N}}$.
(iii): At large RG times $z\gg z_*$ an observer can no longer distinguish the individual objects, and we instead
have the single merged object $\widetilde{\mathcal{O}}^{(\mathcal{N})}$.}
\label{fig:decimation}
\end{figure}

We consider $\mathcal{N}$ a topological operator with support on $Y$.
Partitioning up the Euclidean spacetime into a collection of $D$-dimensional ``pixels'' $P_{i}$, we can
track the effects of RG\ flow (i.e., smearing in the bulk) by increasing the
size of the $P_{i}$ as we move to the infrared. Note that in the linking
between $\mathcal{O}$ and $\mathcal{N}$, a far away observer will only see the action of $\mathcal{N}$ on $\mathcal{O}$, i.e., the
combination $\mathcal{O}^{(\mathcal{N)}}$. In other words, a far away observer cannot
microscopically probe $\mathcal{O}$ and $\mathcal{N}$ separately. This
crossover behavior happens at some scale $z=z_{\ast}$.\footnote{The precise value of $z_{\ast}$ is scheme dependent. Indeed, a
change of RG scheme in the boundary CFT will be reflected as a choice of diffeomorphism in the bulk.
Even so, there is  clearly a characteristic length scale associated with this ``threshold correction'' to the
RG evolution.}
Said differently, for $z\ll z_{\ast}$, the bulk configuration consists of $\widetilde{\mathcal{O}}$, and---provided it exists---will be accompanied by $\widetilde{\mathcal{N}}$, so we denote this by $\widetilde{\mathcal{O}} \oplus \widetilde{\mathcal{N}}$.
For $z\gg z_{\ast}$, we can only detect $\widetilde{\mathcal{O}}^{(\mathcal{N)}}$ due to smearing effects.
See figure \ref{fig:decimation} for a depiction.

Formally speaking, we can also estimate where this crossover takes place. To see how, consider in the CFT\ a spatial $D$-dimensional ball $B$ which includes the $\mathcal{O} \oplus \mathcal{N}$ configuration. We can build a minimal area ``surface'' in the bulk
Euclidean AdS which is homologous to $B$, and which has the same boundary $\partial B$.
This region extends out a finite amount into the bulk,
and this is the demarcation region $z = z_{\ast}$. So,
there is a maximum depth $z_{\ast}$ to which it can penetrate. This is the
crossover scale from the individual objects $\widetilde{\mathcal{O}%
}\mathcal{\oplus}\widetilde{\mathcal{N}}$ to $\widetilde{\mathcal{O}%
}^{(\mathcal{N)}}$. By general scaling arguments, we also know that $z_{\ast} \sim r(B)$, the radius of $B$.

Consider now a field configuration with profile which transitions
from $\widetilde{\mathcal{O}}$ to $\widetilde{\mathcal{O}}^{(\mathcal{N)}}$
and compare this with a field configuration which is simply
$\widetilde{\mathcal{O}}$ throughout (no $\mathcal{N}$ insertion in the
CFT\ dual):%
\begin{equation}\label{eq:versusmode}
\mathcal{(\widetilde{\mathcal{O}}\rightarrow\widetilde{\mathcal{O}%
}^{(\mathcal{N)}})}\text{ \ \ vs \ \ }\mathcal{(\widetilde{\mathcal{O}%
}\rightarrow\widetilde{\mathcal{O}})}\text{.}%
\end{equation}
The two configurations have different bulk stress energy simply because there
is a non-zero jump in the $z$-profile near $z=z_{\ast}$; derivatives $D_z$ along the RG line are different.\footnote{Note that even a jump by a phase rotation is enough to signal a discontinuity in the $z$-direction.} Here, $D_z$ refers to a formal covariant derivative with any possible background connections switched on.

Let us elaborate on this point. Since we expect in an actual AdS/CFT pair that the bulk symmetry will be gauged anyway,
one might ask whether in such a situation the resultant $\widetilde{\mathcal{O}}^{(\mathcal{N})}$ is actually ``gauge equivalent'' to $\widetilde{\mathcal{O}}$.\footnote{We thank the anonymous referee for a helpful question of clarification.} To properly calculate differences in the values of the $\widetilde{\mathcal{O}}$ operator, we introduce a formal Wilson defect operator $\widetilde{\mathcal{W}}(z^{\prime}, z)$ which serves to properly compute differences between charged operators located at different radial slices:
\begin{equation}
\widetilde{\mathcal{O}}(z^{\prime}) \widetilde{\mathcal{W}}(z^{\prime},z) - \widetilde{\mathcal{O}}(z).
\end{equation}
This can be done for any candidate categorical symmetry which has a corresponding Symmetry TFT / Symmetry Theory.
Now, the defect $\widetilde{\mathcal{W}}$ has support on a subspace which has one higher dimension than $\widetilde{\mathcal{O}}$.\footnote{In particular, in the Symmetry TFT / Symmetry Theory this defect would topologically link with $\widetilde{\mathcal{N}}$.} Returning to line (\ref{eq:versusmode}), we can now properly compare the differences:
\begin{equation}
\widetilde{\mathcal{O}}^{(\mathcal{N})}(z^{\prime}) \widetilde{\mathcal{W}}^{(\mathcal{N})}(z^{\prime},z) - \widetilde{\mathcal{O}}(z)
\text{ \ \ vs \ \ }
\widetilde{\mathcal{O}}(z^{\prime}) \widetilde{\mathcal{W}}(z^{\prime},z) - \widetilde{\mathcal{O}}(z)\text{.}%
\end{equation}
For $z^{\prime} = z + \delta z$, these finite differences can be viewed as approximating a covariant derivative $\delta z D_{z} \widetilde{\mathcal{O}}(z)$ associated with the corresponding symmetry. In particular, these differences are what will directly enter into any stress energy tensor.\footnote{Consider for example the case of a complex scalar charged under a $U(1)$ symmetry. Then, the stress energy tensor has contributions from covariant derivatives of the form $T_{ij} \supset D_{i} \phi D_{j} \phi^{\dag}$.} Our smearing argument has established that $\widetilde{\mathcal{O}}^{(\mathcal{N})} \widetilde{\mathcal{W}}^{(\mathcal{N})} \neq \widetilde{\mathcal{O}} \widetilde{\mathcal{W}}$. This establishes the expected jump in the stress energy.

Said differently, this jump in the profile of the bulk $\widetilde{\mathcal{O}}$ before and after $z = z_{\ast}$ allows us to establish two things in one shot. First of all, there must be \textit{something else} present besides just $\widetilde{\mathcal{O}}$ (for $z \ll z_{\ast}$) and $\widetilde{\mathcal{O}}^{(\mathcal{N})}$ (for $z \gg z_{\ast}$). This establishes the existence of $\widetilde{\mathcal{N}}(z)$.

\begin{figure}
\centering
\scalebox{0.8}{
\begin{tikzpicture}
	\begin{pgfonlayer}{nodelayer}
		\node [style=none] (0) at (2.25, 2.5) {};
		\node [style=none] (1) at (2.25, -2.5) {};
		\node [style=none] (2) at (-3.75, 0) {};
		\node [style=none] (5) at (-2, 1.75) {};
		\node [style=none] (6) at (-2, -1.75) {};
		\node [style=none] (8) at (2.25, 0.75) {};
		\node [style=none] (9) at (2.25, -0.75) {};
		\node [style=BrownCircle] (10) at (2.25, 0) {};
		\node [style=none] (11) at (-1.9, 0.9) {$\widetilde{\mathcal{O}}^{(\mathcal{N})}$};
		\node [style=none] (12) at (3, 0) {};
		\node [style=none] (14) at (0, 0.25) {};
		\node [style=none] (15) at (2, 0.15) {};
		\node [style=none] (16) at (2, -0.15) {};
		\node [style=none] (17) at (2.25, 0) {};
		\node [style=none] (18) at (-2, 0.5) {};
		\node [style=none] (20) at (0, 3) {};
		\node [style=none] (21) at (0, -3) {};
		\node [style=none] (24) at (0, -3.5) {$z=z_*$};
		\node [style=none] (25) at (12, 2.5) {};
		\node [style=none] (26) at (12, -2.5) {};
		\node [style=none] (27) at (6, 0) {};
		\node [style=none] (28) at (7.75, 1.75) {};
		\node [style=none] (29) at (7.75, -1.75) {};
		\node [style=BrownCircle] (32) at (12, 0) {};
		\node [style=none] (33) at (7.75, 0.9) {$\widetilde{\mathcal{O}}$};
		\node [style=none] (34) at (12.75, 0) {};
		\node [style=none] (38) at (12, 0) {};
		\node [style=none] (40) at (9.75, 3) {};
		\node [style=none] (41) at (9.75, -3) {};
		\node [style=none] (44) at (9.75, -3.5) {$z=z_*$};
		\node [style=none] (45) at (2.75, 0) {${\mathcal{O}}$};
		\node [style=none] (46) at (2.25, 1.2) {$\mathcal{N}$};
		\node [style=none] (48) at (0, -4.5) {\large (i)};
		\node [style=none] (49) at (9.75, -4.5) {\large (ii)};
		\node [style=none] (50) at (5, -5) {};
		\node [style=none] (51) at (7.75, 0.5) {};
		\node [style=none] (52) at (7.75, -0.5) {};
		\node [style=none] (53) at (-2, -0.5) {};
		\node [style=none] (54) at (0, -0.25) {};
		\node [style=none] (55) at (12, 0.5 ) {$\mathcal{O}$};
		\node [style=none] (55) at (0.5, 0.3) {};
		\node [style=none] (56) at (0.5, -0.3) {};
		\node [style=none] (57) at (1, 0) {};
		\node [style=none] (58) at (0.7, 0.725) {};
		\node [style=none] (59) at (0.7, 0.5) {};
		\node [style=none] (60) at (0.7, -0.5) {};
		\node [style=none] (61) at (0.7, -0.725) {};
		\node [style=none] (62) at (0.85, 1.25) {$\widetilde{\mathcal{O}}\oplus \widetilde{\mathcal{N}}$};
	\end{pgfonlayer}
	\begin{pgfonlayer}{edgelayer}
		\filldraw[fill=purple!40, draw=purple!40]  (-2, 0.5) -- (0, 0.25) -- (0, -0.25) -- (-2, -0.5) -- cycle;
		\filldraw[fill=brown!40, draw=brown!40]  (0, -0.25) -- (0, 0.25) -- (2.25, 0) -- cycle;
		\filldraw[fill=brown!40, draw=brown!40]  (7.75, 0.5) -- (7.75, -0.5) -- (12, 0) -- cycle;
		\filldraw[fill=purple!40, draw=purple!40] (-2, 0) ellipse (0.25cm and 0.5cm);
		\filldraw[fill=brown!40, draw=brown!40] (7.75, 0) ellipse (0.25cm and 0.5cm);
		\draw [style=PurpleLine, bend left=90, looseness=0.55] (8.center) to (9.center);
		\draw [style=ThickLine] (1.center) to (6.center);
		\draw [style=ThickLine] (5.center) to (0.center);
		\draw [style=ThickLine, in=90, out=-165] (5.center) to (2.center);
		\draw [style=ThickLine, in=165, out=-90] (2.center) to (6.center);
		\draw [style=ThickLine, bend right=90, looseness=0.50] (1.center) to (0.center);
		\draw [style=DottedLine, bend right=90, looseness=0.45] (0.center) to (1.center);
		\draw [style=PurpleLine, in=-85, out=-180] (9.center) to (16.center);
		\draw [style=PurpleLine, in=-180, out=85] (15.center) to (8.center);
		\draw [style=BrownLine] (17.center) to (14.center);
		\draw [style=DashedLineThin] (20.center) to (21.center);
		\draw [style=PurpleLine] (14.center) to (18.center);
		\draw [style=ThickLine] (26.center) to (29.center);
		\draw [style=ThickLine] (28.center) to (25.center);
		\draw [style=ThickLine, in=90, out=-165] (28.center) to (27.center);
		\draw [style=ThickLine, in=165, out=-90] (27.center) to (29.center);
		\draw [style=ThickLine, bend right=90, looseness=0.50] (26.center) to (25.center);
		\draw [style=DottedLine, bend right=90, looseness=0.50] (25.center) to (26.center);
		\draw [style=DashedLineThin] (40.center) to (41.center);
		\draw [style=BrownLine, bend left=90, looseness=0.75] (51.center) to (52.center);
		\draw [style=BrownLine, bend right=90, looseness=0.75] (51.center) to (52.center);
		\draw [style=BrownLine] (51.center) to (38.center);
		\draw [style=BrownLine] (38.center) to (52.center);
		\draw [style=PurpleLine, bend right=90, looseness=0.75] (18.center) to (53.center);
		\draw [style=PurpleLine, bend left=90, looseness=0.75] (18.center) to (53.center);
		\draw [style=PurpleLine] (53.center) to (54.center);
		\draw [style=BrownLine] (54.center) to (17.center);
		\draw [style=PPurpleLine, in=90, out=90, looseness=2.75] (55.center) to (57.center);
		\draw [style=PPurpleLine, in=-90, out=-90, looseness=2.75] (57.center) to (56.center);
	\end{pgfonlayer}
\end{tikzpicture}}
\caption{(i): Depiction of a field configuration, as specified by $\widetilde{\mathcal{N}}$, in which $\widetilde{\mathcal{O}}$ transitions to $\widetilde{\mathcal{O}}^{(\mathcal{N})}$. We indicate the symmetry operator $\mathcal{N}$ and charged operator $\mathcal{O}$
of the CFT. When $z\ll z_*$, they smear to $\widetilde{\mathcal{O}} \oplus \widetilde{\mathcal{N}}$. When $z\gg z_*$, they smear to the symmetry transformed operator $\widetilde{\mathcal{O}}^{(\mathcal{N})}$. Near $z=z_*$ we have a discontinuity in the $z$-direction as we jump from $\widetilde{\mathcal{O}}$ to $\widetilde{\mathcal{O}}^{(\mathcal{N})}$. (ii): Depiction of the same setup as in (i) with constant field profile.}
\label{fig:LineChangeOp}
\end{figure}
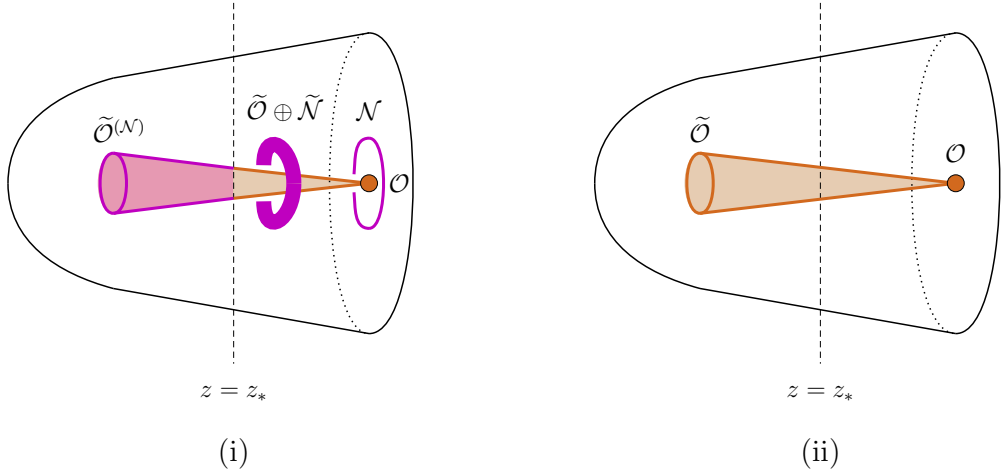

Second, if $\widetilde{\mathcal{N}}$ contributes no stress energy,
then the two configurations $\mathcal{(\widetilde{\mathcal{O}}\rightarrow
\widetilde{\mathcal{O}}^{(\mathcal{N)}})}$
and\ $\mathcal{(\widetilde{\mathcal{O}}\rightarrow\widetilde{\mathcal{O}})}$
would have had the same stress energy, a contradiction. See figure \ref{fig:LineChangeOp}
for a depiction of these two situations.

The resolution of the contradiction is that the insertion of $\widetilde{\mathcal{N}}$ itself
contributes some stress energy to the configuration. Said differently,
returning to line (\ref{Ntildez}), we conclude that there is a non-trivial
coupling of the bulk metric and its local fluctuations to the theory specified by $\widetilde{\mathcal{N}%
}(z)$.

Summarizing, we have argued that coarse graining in the Euclidean CFT leads to a maximal depth of penetration in the
Euclidean AdS, as captured by an associated $D$-dimensional ball of the Euclidean CFT which ``dips'' into the bulk AdS.
One might view this as a Euclidean signature generalization of an RT surface, but it is clear that
this differs from the standard notion which would have made reference to a $(D-1)$-dimensional ball.
Here, we simply view this as a formal device to figure out the crossover region $z = z_{\ast}$. In any event, the existence
of a crossover scale is clear, and this also establishes that $\widetilde{\mathcal{N}}$ exists and
is not topological in the bulk AdS space.

\subsection{Lorentzian Signature Analysis}

We now repeat our coarse graining analysis but in Lorentzian signature. Here, there is an important
distinction in the dimension of support for our $p$-form symmetry operator. Working at a fixed time slice, observe
that for $p = 0$, the corresponding symmetry operator would fill all of space, while for $p \geq 1$
we can put the topological operator on a smaller spatial region. This is unpleasant, and
one workaround is to assume that we can decompose the domain of the TFT into smaller regions,
and analyze the entanglement wedge of each of these objects separately.
This is the approach of references \cite{Harlow:2018jwu, Harlow:2018tng},
and we return to this treatment in section \ref{sec:SPLIT}.

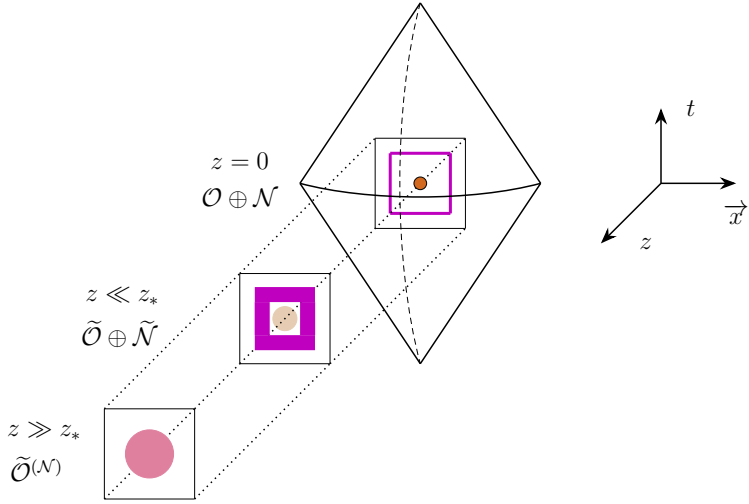
\begin{figure}
\centering
\scalebox{0.8}{
\begin{tikzpicture}
	\begin{pgfonlayer}{nodelayer}
		\node [style=none] (0) at (0, 3) {};
		\node [style=none] (1) at (-2, 0) {};
		\node [style=none] (2) at (2, 0) {};
		\node [style=none] (3) at (0, -3) {};
		\node [style=none] (4) at (-0.5, 0.5) {};
		\node [style=none] (5) at (0.5, 0.5) {};
		\node [style=none] (6) at (0.5, -0.5) {};
		\node [style=none] (7) at (-0.5, -0.5) {};
		\node [style=BrownCircleSmall] (8) at (0, 0) {};
		\node [style=none] (9) at (4, 0) {};
		\node [style=none] (10) at (3, -1) {};
		\node [style=none] (11) at (4, 1.25) {};
		\node [style=none] (12) at (5.25, 0) {};
		\node [style=none] (13) at (3.75, -1) {$z$};
		\node [style=none] (14) at (5.25, -0.5) {$\overrightarrow{x}$};
		\node [style=none] (15) at (4.5, 1.25) {$t$};
		\node [style=none] (16) at (0.75, 0.75) {};
		\node [style=none] (17) at (0.75, -0.75) {};
		\node [style=none] (18) at (-0.75, -0.75) {};
		\node [style=none] (19) at (-0.75, 0.75) {};
		\node [style=none] (20) at (-1.5, -1.5) {};
		\node [style=none] (21) at (-1.5, -3) {};
		\node [style=none] (22) at (-3, -3) {};
		\node [style=none] (23) at (-3, -1.5) {};
		\node [style=none] (24) at (-3.75, -3.75) {};
		\node [style=none] (25) at (-3.75, -5.25) {};
		\node [style=none] (26) at (-5.25, -5.25) {};
		\node [style=none] (27) at (-5.25, -3.75) {};
		\node [style=none] (28) at (-2.75, -1.85) {};
		\node [style=none] (29) at (-1.75, -1.85) {};
		\node [style=none] (30) at (-1.75, -2.65) {};
		\node [style=none] (31) at (-2.75, -2.65) {};
		\node [style=none] (32) at (0, -6) {};
		\node [style=none] (33) at (-2.6275, -1.975) {};
		\node [style=none] (34) at (-1.8725, -1.975) {};
		\node [style=none] (35) at (-1.8725, -2.525) {};
		\node [style=none] (36) at (-2.6275, -2.525) {};
		\node [style=none] (37) at (-4.95, -1.875) {$z\ll z_*$};
		\node [style=none] (38) at (-6.25, -4.125) {$z\gg z_*$};
		\node [style=none] (39) at (-6.375, -4.75) {$\widetilde{\mathcal{O}}^{({\mathcal{N}})}$};
		\node [style=none] (40) at (-5, -2.5) {$\widetilde{\mathcal{O}}\oplus \widetilde{\mathcal{N}}$};
		\node [style=none] (41) at (-3, -0.25) {$\mathcal{O}\oplus\mathcal{N}$};
		\node [style=none] (42) at (-3, 0.375) {$z=0$};
	\end{pgfonlayer}
	\begin{pgfonlayer}{edgelayer}
		\filldraw[fill=brown!40, draw=brown!40] (-2.25, -2.25) ellipse (0.2cm and 0.2cm);
		\draw [style=ThickLine] (1.center) to (0.center);
		\draw [style=ThickLine] (0.center) to (2.center);
		\draw [style=ThickLine] (2.center) to (3.center);
		\draw [style=ThickLine] (3.center) to (1.center);
		\draw [style=PurpleLine] (4.center) to (5.center);
		\draw [style=PurpleLine] (5.center) to (6.center);
		\draw [style=PurpleLine] (6.center) to (7.center);
		\draw [style=PurpleLine] (7.center) to (4.center);
		\draw [style=ThickLine, bend right=15, looseness=0.75] (1.center) to (2.center);
		\draw [style=DashedLineThin, bend right=15, looseness=0.75] (0.center) to (3.center);
		\draw [style=ArrowLineRight] (9.center) to (10.center);
		\draw [style=ArrowLineRight] (9.center) to (12.center);
		\draw [style=ArrowLineRight] (9.center) to (11.center);
		\draw (19.center) to (18.center);
		\draw (18.center) to (17.center);
		\draw (17.center) to (16.center);
		\draw (16.center) to (19.center);
		\draw (23.center) to (22.center);
		\draw (22.center) to (21.center);
		\draw (21.center) to (20.center);
		\draw (20.center) to (23.center);
		\draw [style=DottedLine] (23.center) to (19.center);
		\draw [style=DottedLine] (20.center) to (16.center);
		\draw [style=DottedLine] (22.center) to (20.center);
		\draw [style=DottedLine] (21.center) to (17.center);
		\draw (27.center) to (26.center);
		\draw (26.center) to (25.center);
		\draw (25.center) to (24.center);
		\draw (24.center) to (27.center);
		\draw [style=DottedLine] (26.center) to (24.center);
		\draw [style=DottedLine] (24.center) to (22.center);
		\draw [style=DottedLine] (25.center) to (21.center);
		\draw [style=DottedLine] (27.center) to (23.center);
		\draw [style=SBrownLine] (28.center) to (29.center);
		\draw [style=SBrownLine] (30.center) to (31.center);
		\draw [style=SBrownLine] (33.center) to (36.center);
		\draw [style=SBrownLine] (34.center) to (35.center);
         \filldraw[fill=purple!50, draw=purple!50] (-4.5, -4.5) ellipse (0.4cm and 0.4cm);
	\end{pgfonlayer}
\end{tikzpicture}}
\caption{Depiction of smearing for the $\mathcal{O} \oplus \mathcal{N}$ configuration of the boundary CFT. In Lorentzian signature the subspace filled by $\mathcal{N}$ sweeps out both a spatial and temporal directions. Surrounding the configuration with a causal diamond in the CFT, we can track the associated entanglement wedge bounded by the HRT surface. This leads to a maximal depth for disambiguating the two constituents; beyond this depth we instead have a single $\widetilde{\mathcal{O}}^{(\mathcal{N})}$ in the bulk. This again leads to a contradiction unless the bulk $\widetilde{\mathcal{N}}$ is non-topological.}
\label{fig:Diamond}
\end{figure}

Here, we shall instead take a different tack: to keep our treatment of all $p$-form symmetries
on an equal footing, we consider operators $\mathcal{O}$ supported on a spatial subspace, and so our $\widetilde{\mathcal{N}}$ will necessarily have some finite extent in both space and time. For example, in the case of a 2D Lorentzian signature CFT we can speak of a local operator $\mathcal{O}(x)$ which links with $\mathcal{\mathcal{N}}$, supported on a closed one-dimensional curve which
has both spacelike and timelike pieces. See figure \ref{fig:Diamond} for a depiction.

In this case, we can draw a large causal diamond $R$ around the $\mathcal{O} \oplus \mathcal{N}$ configuration. On short distance scales, we can
again resolve the two individual constituents, but at long distance scales we instead only detect $\mathcal{O}^{(\mathcal{N})}$.
This is apparent in the holographic dual by constructing the HRT surface $EW(R)$ associated to our region $R$. Inside $EW(R)$, we can resolve $\widetilde{\mathcal{O}} \oplus \widetilde{\mathcal{N}}$, but outside, the two constituents have merged to $\widetilde{\mathcal{O}}^{(\mathcal{N})}$. Now everything proceeds as before; we get a maximum depth of resolution set by a characteristic scale $z_{\ast}$, and again, the ``jump'' in comparing the stress energy of the $(\widetilde{\mathcal{O}} \rightarrow \widetilde{\mathcal{O}})$ line and the $(\widetilde{\mathcal{O}} \rightarrow \widetilde{\mathcal{O}}^{(\mathcal{N})})$ line tells us that $\widetilde{\mathcal{N}}$ exists and cannot be purely topological in the bulk. See figure \ref{fig:Diamond} for a depiction of the causal diamond in the boundary CFT, as well as the smearing in the bulk.

As a final comment, note that if we had restricted to $p \geq 1$ with a topological operator supported on a purely spatial region
then we could have run precisely the same coarse graining argument used
in our Euclidean signature analysis, but now for just spatial subregions of the CFT.

\subsection{$(-1)$-form Symmetries}

We now return to the case of $(-1)$-form symmetries of the boundary CFT.
Our aim here will be to give a bottom up proposal for how to make sense of
this case in a way compatible with holographic considerations.

Recall that $(-1)$-form symmetries are associated with specifying the
parameters of the CFT, which we label as $\{\lambda\}$. The associated
topological operator $\mathcal{N}$ fills the spacetime so the notion of
\textquotedblleft smearing\textquotedblright\ itself would appear to be
somewhat ill-defined. Nevertheless, we can construct a space $\Lambda$ from the $\{\lambda\}$ and equip it with
the Zamolodchikov metric \cite{Zamolodchikov:1986gt} for the continuous
parameters and the discrete topology for any discrete parameters. From this
perspective, we get a family of CFTs fibered over $\Lambda$, and so we can
speak of a spacetime filling topological operator $\mathcal{N}(\lambda)$ which
sits at a particular point $\lambda\in\Lambda$ and fills the directions of the
CFT. Inserting $\mathcal{N}(\lambda)$ moves us from the point $\lambda$ to
the point $\lambda^{(\mathcal{N(\lambda)})}$.

What is the holographic dual $\widetilde{\mathcal{N}}(\lambda,z)$?
To begin, recall that via the standard holographic dictionary, the parameter
$\lambda$ is to be viewed as the asymptotic profile of the non-normalizable
component of a bulk field $\phi$ with $\phi|_{\partial AdS}\sim\lambda$
\cite{Witten:1998qj, Gubser:1998bc}. Suppose that we now pull $\mathcal{N}%
(\lambda)$ off the boundary CFT located at $z=0$. We now have a bulk insertion
of a codimension one object (i.e., a ``wall''). Crossing from one side to the other modifies the value of
$\phi$, i.e., it induces a jump in the bulk modulus field.

As far as establishing that $\widetilde{\mathcal{N}}(\lambda_{1}%
\rightarrow\lambda_{2})$ is not topological in the bulk, the argument proceeds
much as in the previous cases, i.e., we proceed via proof by contradiction. We
observe that the bulk modulus field \textquotedblleft
charged\textquotedblright\ under $\widetilde{\mathcal{N}}(\lambda
_{1}\rightarrow\lambda_{2})$ jumps, and so the presence of a kink profile
would induce a non-zero stress energy. There is no contradiction provided
$\widetilde{\mathcal{N}}(\lambda_{1}\rightarrow\lambda_{2})$ supports a
non-topological worldvolume action in the bulk.

It is also of interest to consider possible wormhole configurations which
connect different values of $\lambda$, as in references \cite{Witten:1999xp,
Maldacena:2004rf}. For two values of parameters $\lambda_{1}$ and $\lambda
_{2}$ in the CFT, suppose we have a $(-1)$-form symmetry operator
$\mathcal{N}(\lambda_{1}\rightarrow\lambda_{2})$ which connects the two. In
the gravity dual, we have an interpolating bulk geometry with a
$\widetilde{\mathcal{N}}(\lambda_{1}\rightarrow\lambda_{2})$ wall inserted.
Observe that this insertion is orientable, and so $\widetilde{\mathcal{N}%
}(\lambda_{1}\rightarrow\lambda_{2})^{\dag}$ implements the opposite
transformation.\footnote{If $\mathcal{N}(\lambda_{1}\rightarrow\lambda_{2})$
is non-invertible, this can lead to some hysteresis.}

\begin{figure}[t!]
\centering
\scalebox{0.7}{\begin{tikzpicture}
	\begin{pgfonlayer}{nodelayer}
		\node [style=none] (0) at (-2.75, 6.5) {};
		\node [style=none] (1) at (-2.75, 1.5) {};
		\node [style=none] (2) at (-8.75, 4) {};
		\node [style=none] (3) at (-7, 5.75) {};
		\node [style=none] (4) at (-7, 2.25) {};
		\node [style=none] (32) at (-5, -7.25) {(ii)};
		\node [style=none] (33) at (4, 0.5) {(iii)};
		\node [style=none] (34) at (0, -8) {};
		\node [style=none] (40) at (-2.75, -1) {};
		\node [style=none] (41) at (-2.75, -6) {};
		\node [style=none] (42) at (-8.75, -3.5) {};
		\node [style=none] (43) at (-7, -1.75) {};
		\node [style=none] (44) at (-7, -5.25) {};
		\node [style=none] (45) at (-5.75, -1.52) {};
		\node [style=none] (46) at (-5.75, -5.475) {};
		\node [style=none] (47) at (-5, 0.5) {(i)};
		\node [style=none] (60) at (-5.75, -3.5) {$\widetilde{\mathcal{N}}$};
		\node [style=none] (62) at (-2.75, 4) {${\mathcal{N}}$};
		\node [style=none] (63) at (0, 2) {};
		\node [style=none] (64) at (4, 5) {};
		\node [style=none] (65) at (6, 2) {};
		\node [style=none] (66) at (10, 5) {};
		\node [style=NodeCross] (67) at (3.75, 3.5) {};
		\node [style=NodeCross] (68) at (6.25, 3.5) {};
		\node [style=none] (69) at (3.75, 3.75) {};
		\node [style=none] (70) at (3.75, 6.5) {};
		\node [style=none] (71) at (6.25, 3.75) {};
		\node [style=none] (72) at (6.25, 6.5) {};
		\node [style=none] (73) at (3.75, 6.75) {CFT$_\lambda$};
		\node [style=none] (74) at (6.625, 6.75) {CFT$_{\lambda^{(\mathcal{N}(\lambda))}}$};
		\node [style=none] (75) at (4.25, 3.5) {};
		\node [style=none] (76) at (5.75, 3.5) {};
		\node [style=none] (77) at (5, 4) {${\mathcal{N}}$};
		\node [style=none] (78) at (9, 2.75) {$\Lambda=\{\lambda \}$};
		\node [style=none] (79) at (6.5, -0.5) {};
		\node [style=none] (80) at (6.5, -3) {};
		\node [style=none] (81) at (6.5, -4) {};
		\node [style=none] (82) at (6.5, -6.5) {};
		\node [style=none] (84) at (8, -1.75) {CFT$_\lambda$};
		\node [style=none] (85) at (8.375, -5.25) {CFT$_{\lambda^{(\mathcal{N}(\lambda))}}$};
		\node [style=none] (86) at (1, -3.5) {};
		\node [style=none] (87) at (2.75, -3.5) {};
		\node [style=none] (88) at (0.25, -3.5) {$\widetilde{\mathcal{N}}$};
		\node [style=none] (89) at (4, -7.25) {(iv)};
		\node [style=none] (91) at (-2.75, 1) {CFT$_{\lambda^{(\mathcal{N}(\lambda))}}$};
		\node [style=none] (92) at (-2.75, -6.5) {CFT$_\lambda$};
	\end{pgfonlayer}
	\begin{pgfonlayer}{edgelayer}
		\filldraw[fill=purple!40, draw=purple!40] (-5.75, -3.5) ellipse (0.59cm and 1.98cm);
		\filldraw[fill=purple!70, draw=purple!70] (-2.75, 4) ellipse (0.72cm and 2.5cm);
		\filldraw[fill=purple!40, draw=purple!40] (1.875, -3.5) ellipse (0.9cm and 0.25cm);
		\draw [style=ThickLine] (1.center) to (4.center);
		\draw [style=ThickLine] (3.center) to (0.center);
		\draw [style=ThickLine, in=90, out=-165] (3.center) to (2.center);
		\draw [style=ThickLine, in=165, out=-90] (2.center) to (4.center);
		\draw [style=ThickLine, bend right=90, looseness=0.50] (1.center) to (0.center);
		\draw [style=DottedLine, bend right=90, looseness=0.50] (0.center) to (1.center);
		\draw [style=ThickLine] (41.center) to (44.center);
		\draw [style=ThickLine] (43.center) to (40.center);
		\draw [style=ThickLine, in=90, out=-165] (43.center) to (42.center);
		\draw [style=ThickLine, in=165, out=-90] (42.center) to (44.center);
		\draw [style=ThickLine, bend right=90, looseness=0.50] (41.center) to (40.center);
		\draw [style=ThickLine, bend right=90, looseness=0.50] (40.center) to (41.center);
		\draw [style=ThickLine, bend right=90, looseness=0.50] (46.center) to (45.center);
		\draw [style=ThickLine, bend right=90, looseness=0.50] (45.center) to (46.center);
		\draw [style=ThickLine] (63.center) to (64.center);
		\draw [style=ThickLine] (64.center) to (66.center);
		\draw [style=ThickLine] (66.center) to (65.center);
		\draw [style=ThickLine] (65.center) to (63.center);
		\draw [style=DottedLine] (70.center) to (69.center);
		\draw [style=DottedLine] (72.center) to (71.center);
		\draw [style=ArrowLineRight] (75.center) to (76.center);
		\draw [style=ThickLine, bend right=90, looseness=0.50] (80.center) to (79.center);
		\draw [style=ThickLine, bend right=90, looseness=0.50] (79.center) to (80.center);
		\draw [style=ThickLine, bend right=90, looseness=0.50] (82.center) to (81.center);
		\draw [style=ThickLine, bend right=90, looseness=0.50] (81.center) to (82.center);
		\draw [style=ThickLine, bend right=90, looseness=0.50] (87.center) to (86.center);
		\draw [style=ThickLine, bend right=90, looseness=0.50] (86.center) to (87.center);
		\draw [style=ThickLine, in=180, out=90] (87.center) to (80.center);
		\draw [style=ThickLine, in=-90, out=-180] (81.center) to (87.center);
		\draw [style=ThickLine, in=-180, out=90] (86.center) to (79.center);
		\draw [style=ThickLine, in=-180, out=-90] (86.center) to (82.center);
	\end{pgfonlayer}
\end{tikzpicture}}
\caption{(i): Sketch of a CFT with parameter $\lambda^{(\mathcal{N}(\lambda))}$ as a CFT with parameter $\lambda$ stacked with the spacetime filling operator $\mathcal{N}$. (ii): Deformation of $\mathcal{N}$ into the bulk giving $\widetilde{\mathcal{N}}$. (iii): Stackings of $\mathcal{N}$ are interpreted as transformations in the parameter space $\Lambda$. (iv): Wormholes between asymptotically AdS spaces with asymptotic CFTs located at different points in $\Lambda$ contain a codimension one wall set by $\widetilde{\mathcal{N}}$.}
\label{fig:WH}
\end{figure}
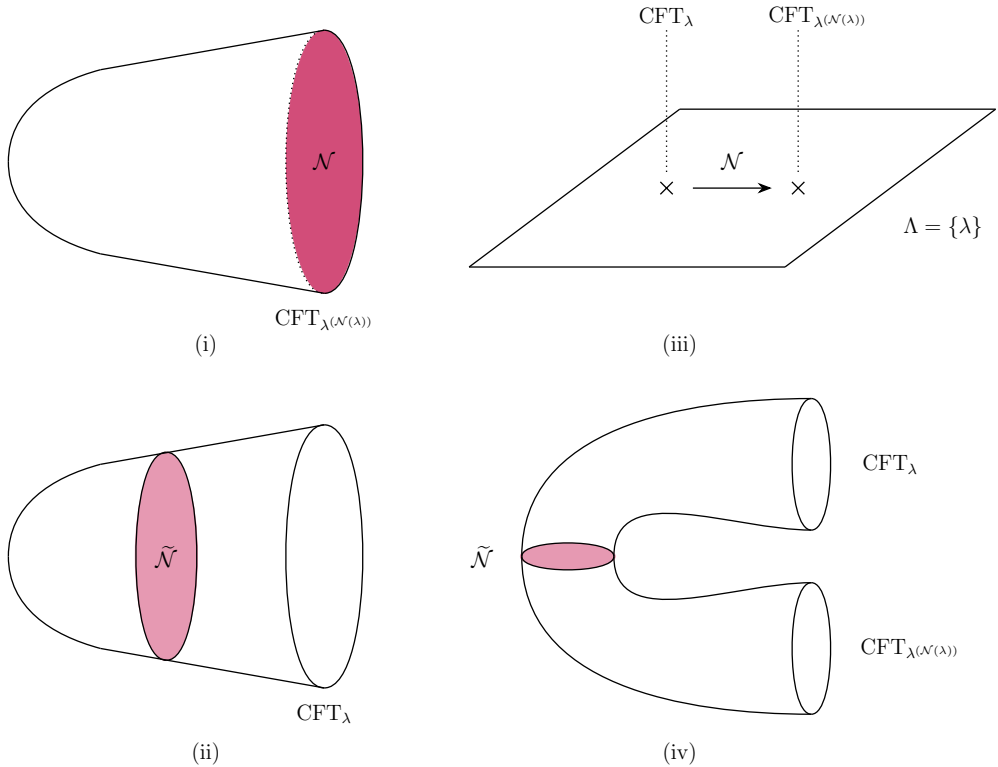

See figure \ref{fig:WH} for a depiction of these bulk configurations with a wall.

\subsection{SymTFT\ Considerations}

Summarizing, we have shown that in the context of the AdS/CFT\ correspondence,
any putative topological global symmetry operator of the boundary CFT\ becomes
non-topological in the bulk. Conversely, we have also shown that in the bulk, there are no
candidate topological symmetry operators for a $p$-form symmetry for $p \geq 0$,
thus excluding possible global symmetries in the gravity dual.

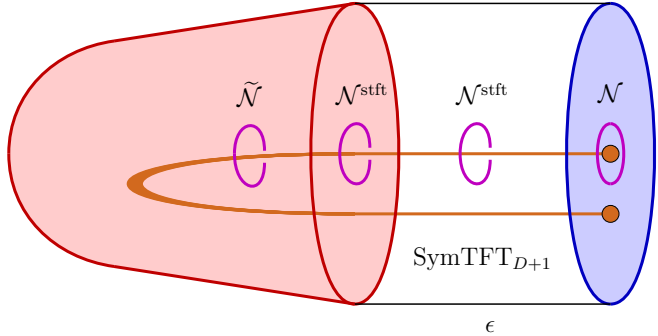
\begin{figure}
\centering
\scalebox{0.8}{
\begin{tikzpicture}
	\begin{pgfonlayer}{nodelayer}
		\node [style=none] (0) at (6.75, 3.5) {};
		\node [style=none] (1) at (6.75, -1.5) {};
		\node [style=none] (2) at (1, 1) {};
		\node [style=none] (6) at (6, -1.5) {};
		\node [style=none] (8) at (2.75, 2.875) {};
		\node [style=none] (9) at (2.75, -0.875) {};
		\node [style=none] (10) at (2.75, 1) {};
		\node [style=none] (11) at (11, 3.5) {};
		\node [style=none] (12) at (11, -1.5) {};
		\node [style=none] (13) at (8.875, 1.5) {};
		\node [style=none] (14) at (8.875, 0.5) {};
		\node [style=none] (15) at (8.375, 1.5) {};
		\node [style=none] (16) at (7.7, 1.5) {};
		\node [style=none] (17) at (6.75, 1.5) {};
		\node [style=none] (18) at (6.75, 0.5) {};
		\node [style=none] (19) at (4.5, 1.5) {};
		\node [style=none] (20) at (4.5, 0.5) {};
		\node [style=none] (21) at (6, 1.5) {};
		\node [style=none] (22) at (6, 0.5) {};
		\node [style=none] (24) at (8.875, -0.75) {SymTFT$_{D+1}$};
		\node [style=none] (25) at (8.875, 2) {$\mathcal{N}^{\text{stft}}$};
		\node [style=none] (26) at (8, -2.25) {};
		\node [style=none] (27) at (11, 1.5) {};
		\node [style=none] (28) at (11, 0.5) {};
		\node [style=none] (29) at (11, 2) {$\mathcal{N}$};
		\node [style=none] (30) at (5, 2) {$\widetilde{\mathcal{N}}$};
		\node [style=none] (32) at (9, -1.875) {$\epsilon$};
        \node [style=none] (33) at (10.035, 1.5) {};
		\node [style=none] (34) at (9.3, 1.5) {};
        \node [style=none] (35) at (6.875, 2) {$\mathcal{N}^{\text{stft}}$};
        \node [style=none] (36) at (5.775, 1.5) {};
		\node [style=none] (37) at (4.9, 1.5) {};
		\node [style=none] (38) at (11, 1) {};
		\node [style=none] (39) at (6.75, 1) {};
		\node [style=none] (40) at (6.75, 0) {};
		\node [style=none] (41) at (11, 0) {};
		\node [style=none] (42) at (9, 1.1) {};
		\node [style=none] (43) at (9, 0.9) {};
		\node [style=none] (44) at (8.5, 1) {};
		\node [style=none] (45) at (5.25, 1.05) {};
		\node [style=none] (46) at (5.25, 0.85) {};
		\node [style=none] (47) at (4.75, 1) {};
		\node [style=none] (48) at (7, 1.1) {};
		\node [style=none] (49) at (7, 0.9) {};
		\node [style=none] (50) at (6.5, 1) {};
  \node [style=BrownCircle] (54) at (11, 1) {};
   \node [style=BrownCircle] (55) at (11, 0) {};
	\end{pgfonlayer}
	\begin{pgfonlayer}{edgelayer}
		\filldraw[fill=red!20, draw=red!20] (2.85, 1) ellipse (1.82cm and 1.85cm);
		\filldraw[fill=red!20, draw=red!20] (6.75, 1) ellipse (0.75cm and 2.5cm);
		\filldraw[fill=blue!20, draw=blue!20] (11, 1) ellipse (0.75cm and 2.5cm);
		\filldraw[fill=red!20, draw=red!20]  (2.75, 2.875) -- (2.75, -0.875) -- (6.75, -1.5) -- (6.75, 3.5) -- cycle;
		\draw [style=BrownLine] (38.center) to (39.center);
		\draw [style=BrownLine, bend right=90, looseness=12.625] (39.center) to (40.center);
		\draw [style=BrownLine] (40.center) to (41.center);
		\draw [style=BrownLine, bend right=90, looseness=12.375] (39.center) to (40.center);
		\draw [style=BrownLine, bend right=90, looseness=12.125] (39.center) to (40.center);
		\draw [style=BrownLine, bend right=90, looseness=12.875] (39.center) to (40.center);
		\draw [style=BrownLine, bend right=90, looseness=12.50] (39.center) to (40.center);
		\draw [style=BrownLine, bend right=90, looseness=12.25] (39.center) to (40.center);
		\draw [style=BrownLine, bend right=90, looseness=12.75] (39.center) to (40.center);
		\draw [style=RedLine] (1.center) to (9.center);
		\draw [style=RedLine, in=-90, out=172] (9.center) to (2.center);
		\draw [style=RedLine, in=90, out=-172] (8.center) to (2.center);
		\draw [style=RedLine] (8.center) to (0.center);
		\draw [style=RedLine, bend left=90, looseness=0.50] (0.center) to (1.center);
		\draw [style=RedLine, bend right=90, looseness=0.50] (0.center) to (1.center);
		\draw [style=BlueLine, bend left=90, looseness=0.50] (11.center) to (12.center);
		\draw [style=BlueLine, bend right=90, looseness=0.50] (11.center) to (12.center);
		\draw [style=ThickLine] (0.center) to (11.center);
		\draw [style=ThickLine] (1.center) to (12.center);
		\draw [style=PurpleLine, bend left=90, looseness=0.75] (27.center) to (28.center);
		\draw [style=PurpleLine, bend right=90, looseness=0.75] (27.center) to (28.center);
 		\draw [style=PurpleLine, in=90, out=90, looseness=3.00] (42.center) to (44.center);
		\draw [style=PurpleLine, in=-90, out=-90, looseness=3.00] (43.center) to (44.center);
		
		\draw [style=PurpleLine, in=90, out=90, looseness=3.00] (45.center) to (47.center);
		\draw [style=PurpleLine, in=-90, out=-90, looseness=3.00] (46.center) to (47.center);
		
		\draw [style=PurpleLine, in=90, out=90, looseness=3.00] (48.center) to (50.center);
		\draw [style=PurpleLine, in=-90, out=-90, looseness=3.00] (49.center) to (50.center);
		
	\end{pgfonlayer}
\end{tikzpicture}}
\caption{Sketch of symmetry operators (purple) in the holographic setup linking with defect operators (brown). The topological symmetry operator $\mathcal{N}$ is localized at the boundary CFT. This operator can be pushed into the bulk of the SymTFT$_{D+1}$ which is now viewed as a subsector of the bulk gravitational dual.}
\label{fig:SymOps}
\end{figure}

This is in accord with what we expect to happen in any $D$-dimensional
QFT\ with global categorical symmetries. From this broader perspective, one
can consider a $(D+1)$-dimensional topological field theory SymTFT$_{D+1}$
which captures these global symmetries. Heavy defects
of the QFT$_{D}$ are extended by one dimension in the bulk, and the
topological operators remain of the same dimension so that linking is
maintained. This naturally suggests that the SymTFT\ is simply a topological
subsector of the AdS/CFT\ correspondence, and this is indeed how it arises in
all known stringy realizations of SymTFTs \cite{Aharony:1998qu,
Heckman:2017uxe, Apruzzi:2021nmk, Baume:2023kkf}. This extension by one
further dimension indicates that the $p$-form symmetries extends to a bulk
gauge field $A_{p+1}$ which produces an extended defect operator,
with the boundary operator attached to the end (see figure \ref{fig:SymOps}).

The SymTFT$_{D+1}$ should be viewed as the low energy limit of a gravitational system.
In terms of embedding it in the AdS/CFT correspondence, one can view the physical boundary condition as being smeared out over much of the
$D+1$ dimensions, with a small purely topological sliver, as in figure \ref{fig:SymOps}.
The minimal way to incorporate gravity is to give each of the fields of the TFT\ some
non-zero kinetic term. Note that this makes sense even for torsional fields
since we can instead work in terms of continuous valued forms, reaching the
formulation in terms of torsional fields by performing a suitable
rescaling.\footnote{As an example, consider the 4D $BF$ theory with topological
action $\frac{k}{2\pi}\int B\wedge F$. In this formulation $B$ and $F$ are
$U(1)$ valued. We can rescale each field by $2\pi/k$ to instead work in terms
of torsional valued fields with action $\frac{2\pi}{k}\int b\wedge f$, in the
obvious notation.}

\subsection{Comments on Topological vs Non-Topological}

The crux of the argument presented above is that the worldvolume action used to define
$\widetilde{\mathcal{N}}$ must have non-trivial metric dependence because otherwise we would observe a jump in the stress energy between $\widetilde{\mathcal{O}}$ and $\widetilde{\mathcal{O}}^{(\mathcal{N})}$. On the other hand,
one might ask what role gravity played in these considerations. Said differently, can there be such jumps induced in a QFT decoupled from gravity and if so, is the line of reasoning just presented truly valid?

To illustrate, consider a $0$-form symmetry operator $\mathcal{N}$ of a CFT$_D$ which admits non-trivial line operators which are charged under this symmetry. This can happen, especially in theories with a 2-group symmetry where a line can be charged under both a 1-form and 0-form symmetry. Now, suppose we take a line operator $\mathcal{L}$ which pierces through the codimension one wall defined by the $0$-form symmetry. On one side we have $\mathcal{L}$, while on the other side we have $\mathcal{L}^{(\mathcal{N})}$. There is a jump in the operator in this case as well. This looks similar to the case just considered, so it is natural to ask whether we are right in arguing for the non-topological nature of $\widetilde{\mathcal{N}}$.

The phenomenon of line-changing operators is rather commonplace, and can occur in both CFTs as well as TFTs. For some discussion of the relation between such effects and 2-groups, see for example \cite{Bhardwaj:2021wif, Lee:2021crt}. In all of these cases, the jump in the line is encapsulated in terms of a defect operator $\mathcal{D}$ localized at the transition point. The defect $\mathcal{D}$ is typically also topological, and this can also be verified in terms of the stringy ``branes at infinity'' picture \cite{Apruzzi:2022rei, GarciaEtxebarria:2022vzq, Heckman:2022muc}. The point here is that in the QFT case, we clearly see some additional degree of freedom has been explicitly inserted near the jumping point.\footnote{For example, consider a stack of D3-branes filling $\mathbb{R}^{3,1}$ and sitting at a point of a transverse $\mathbb{C}^3$. We get duality defects by wrapping constant axio-dilaton 7-branes on the boundary $S^5$ ``at infinity,'' as in \cite{Heckman:2022xgu}. We get a line defect via an F1 / D1 string (the choice depends on the polarization) which stretches from the origin out to the boundary $S^5$. Consider a line operator which crosses from one side of the duality wall to the other. At the wall crossing, there is a light degree of freedom stretching from the string to the 7-brane, but this is localized at the boundary $S^5$, far from the D3-brane stack. This is the stringy implementation of $\mathcal{D}$.}

The most important difference in comparing our AdS analysis with this QFT example is that in the latter we clearly see an extra defect has been included at the transition point; in the AdS case, however, there is a transition from a specific field profile from $\widetilde{\mathcal{O}}$ to $\widetilde{\mathcal{O}}^{(\mathcal{N})}$ with seemingly no such accompanying defect. The absence of a new degree of freedom (topological or otherwise) produces a more dramatic discontinuity, and as we argued previously, this can be resolved by assuming that $\widetilde{\mathcal{N}}(z)$ for $z \neq 0$ also contributes to the bulk stress energy. Indeed, along each $z$-slicing we have no such collision between symmetry defects and CFT operators in the first place; the $\widetilde{\mathcal{N}}$ has simply dissolved into $\widetilde{\mathcal{O}}$ for $z > z_{\ast}$. This is qualitatively different from the way line-changing operators arise in the QFT with gravity decoupled.

\subsection{Properties of $\widetilde{\mathcal{N}}$}

With these observations in place, we now argue that we can indeed interpret
the worldvolume action for $\widetilde{\mathcal{N}}$ as that of a dynamical brane.
To frame the discussion to follow, it is helpful to note that in stringy constructions of generalized symmetry
operators, there is a natural expectation that many examples will result from
suitably wrapped branes in backgrounds of the form $AdS\times X$. In this
context, the worldvolume theory in the AdS directions is given by a suitable
dimensional reduction of the DBI\ and WZ\ terms supported on the D-brane. Pushing the brane to the
conformal boundary of the AdS freezes out the metric dependent terms, resulting in a purely topological action
in the CFT \cite{Apruzzi:2022rei} (see also \cite{GarciaEtxebarria:2022vzq, Heckman:2022muc}).
Of course, one can also contemplate more general solitonic objects, but this
provides a natural first class of examples.

In all of these examples, the brane in question will source some amount of
stress energy. For a gravitational brane wrapped on a $q$-dimensional subspace
$Y$, we can parameterize our ignorance of the brane dynamics by introducing an
unknown worldvolume tensor $M_{ij}(g_{AB},...)$ which depends on the pullback of the bulk
metric $g_{AB}$, and possibly other bulk and worldvolume degrees of freedom on the
brane:
\begin{equation}
S_{\text{brane}}\sim\tau_{q}\underset{Y}{\int}d^{q}\xi\text{ }\sqrt{\det
M}+\mathrm{Topological \, Terms},
\end{equation}
with $\tau_{q}$ an overall dimensionful constant which we refer to as the
tension. For example, in the special case of a D-brane,
we would write $M_{ij}=h_{ij} + b_{ij} + 2\pi\alpha^{\prime} F_{ij}+...$,
with $h_{ij}$ the pullback of the bulk metric to the brane, $b_{ij}$ the
pullback of the NS 2-form, and $F_{ij}$ the field strength tensor of a $U(1)$
gauge field. In this case, the ``Topological Terms'' are just the WZ terms.

What can we say about the value of $\tau_{q}$ in our case? Without providing
further details on the precise AdS/CFT\ pair, we should likely only expect to
obtain crude estimates. That being said, general bottom up considerations
provide some insight.

On dimensional analysis grounds, we know that $\tau_{q}$ ought to scale as:%
\begin{equation}
\tau_{q}\sim\frac{1}{\ell_{\ast}^{q}}.
\end{equation}
for some characteristic length $\ell_{\ast}$. We view this as specifying the
Compton wavelength for our object $\widetilde{\mathcal{N}}$. Now, if
$\widetilde{\mathcal{N}}$ were of size the AdS radius there would be no sense
in which we could localize it in a small region near the boundary of the CFT;
it would immediately be spread over the entirety of our spacetime. Since the
coarse graining takes some RG time to proceed, we conclude that the radius of
curvature $\ell_{\text{AdS}}\gg\ell_{\ast}$. On the other hand, we also know
that this characteristic scale must be greater than the $(D+1)$-dimensional
Planck length $\ell_{\text{pl}}$; otherwise there would be no regime of
validity to consider this object in the effective field theory in the first
place. In principle, however, it could be separated from the Planck scale, and
this possibility does occur in many situations.

Putting these observations together, we learn:%
\begin{equation}
\ell_{\text{AdS}}\gg\ell_{\ast} \gtrsim \ell_{\text{pl}}\text{,}%
\end{equation}
which is not altogether surprising. We comment here that in the special case of a
D-brane action we would of course interpret $\ell_{\ast}$ as the minimal
resolving power of a D-brane \cite{Shenker:1995xq}.

\section{Examples}

In this section we briefly discuss some examples illustrating the general structure found above.\footnote{We thank M. Montero and H. Ooguri
for several questions in this direction.}
In most cases, we can directly deduce the topological sector of $\widetilde{\mathcal{N}}$ by appealing to
its construction in the SymTFT$_{D+1}$. Many examples with holographic duals
have also been constructed from a ``top down'' point of view, see e.g., references \cite{Apruzzi:2022rei,
GarciaEtxebarria:2022vzq, Heckman:2022muc, Heckman:2022xgu, Dierigl:2023jdp,
Cvetic:2023plv, Etheredge:2023ler, Lawrie:2023tdz, Bah:2023ymy, Apruzzi:2023uma, Cvetic:2023pgm, Yu:2023nyn, Gould:2023wgl}.
In most of these cases one considers a brane wrapped ``at infinity'' (prior to taking the near horizon limit) which nonetheless still links with the defect in question. In these cases the worldvolume theory for the $\widetilde{\mathcal{N}}$ is simply that of the corresponding brane. This provides a systematic way to construct symmetry operators (and their categorical generalizations) for various discrete and continuous symmetries.

On the other hand, it is natural to ask whether we need the full machinery of string theory to identify more detailed properties of $\widetilde{\mathcal{N}}$. Here we consider a few examples of this sort, focusing on the symmetry operators for some continuous $0$-form symmetries. We first discuss in detail the case of a $U(1)$ symmetry and then turn to $G$ a continuous Lie group with a single connected component.

Consider, then, a CFT$_D$ with a $U(1)$ $0$-form symmetry which is dual to a $U(1)$ gauge symmetry in the bulk. Since we have a conserved current in the boundary theory, there is a current $(D-1)$-form $j_{D-1}$ which we can integrate over a $(D-1)$-dimensional closed subspace $Y_{D-1}$ to form a symmetry operator:
\begin{equation}
\mathcal{N}_{\eta} = \exp \left(2 \pi i \eta \underset{Y_{D-1}}{\int} j_{D-1} \right).
\end{equation}
This links with local operators which are charged under the $U(1)$ global symmetry.
In the CFT$_D$ we can introduce a background gauge field $a_1$ associated with this global
symmetry.

Now, in the bulk AdS$_{D+1}$ we have a $U(1)$ gauge field $A_1$
which approaches the background value $a_1$ at the boundary.
Denote by $\mathcal{W}$ an electric line operator built from $A_1$.
Let us consider the idealized situation where gravity is switched off (as in the SymTFT$_{D+1}$). Then, we can construct a 1-form symmetry operator which links with $\mathcal{W}$:
\begin{equation}\label{eq:NprimeU(1)}
\mathcal{N}^{\text{stft}}_{\eta}(z) = \exp \left(2 \pi i \eta \underset{Y_{D-1}(z)}{\int} F^{\text{dual}} \right),
\end{equation}
where in the above we have assumed that the operator is close enough to the conformal boundary
that we can take $Y_{D-1}(z)$ homotopic to $Y_{D-1}$.

What happens as we take $z \rightarrow 0$? Since it is topological, we can push
$\mathcal{N}^{\text{stft}}_{\eta}(z)$ all the way to the boundary. Doing so, we get:
\begin{equation}
\mathcal{N}^{\text{stft}}_{\eta}(z = 0) = \exp \left(2 \pi i \eta \underset{Y_{D-1}}{\int} F^{\text{dual}} \right).
\end{equation}
Next, introduce a $D$-chain $C_D$ which terminates in AdS$_{D+1}$ and has boundary $\partial C_{D} = Y_{D-1}$.
Then, we can equivalently write:
\begin{equation}
\mathcal{N}^{\text{stft}}_{\eta}(z = 0) = \exp \left(2 \pi i \eta \underset{C_{D}}{\int} d F^{\text{dual}} \right).
\end{equation}
On the other hand, we also know, via the equations of motion, that
\begin{equation}
d F^{\text{dual}} = J_{D}.
\end{equation}
As such, we also have:
\begin{equation}
\mathcal{N}^{\text{stft}}_{\eta}(z = 0) = \exp \left(2 \pi i \eta \underset{C_{D}}{\int} J_{D} \right) = \exp \left(2 \pi i \eta \underset{Y_{D-1}}{\int} j_{D-1} \right),
\end{equation}
where in the last equality we used the fact that the topological linking in the CFT$_{D}$
can equivalently be computed in terms of an intersection number in the bulk AdS$_{D+1}$.\footnote{In slightly more detail, we can view the electric line operator as an insertion in the action along the
Poincar\'{e} dual (PD) of $J_D$, namely $\delta_{PD(J_D)}$. The integral over $J_D$ thus collapses to the intersection between the supports of $C_D$ and $PD(J_D)$.} Summarizing, we have found that in the bulk, the topological part of $\widetilde{\mathcal{N}}(z)$ is obtained
via line (\ref{eq:NprimeU(1)}).

Once gravity is switched on there will be additional non-topological contributions to the worldvolume,
as captured by $\widetilde{\mathcal{N}}$ rather than $\mathcal{N}^{\text{stft}}$, much as in figure \ref{fig:SymOps}.
What can we say about these non-topological terms? As mentioned earlier, this is a model dependent issue but
we can still deduce the general form to be a DBI-like action in many cases of interest. Suppose, for example, that our
AdS$_{D+1}$/CFT$_{D}$ pair arises from a geometry of the form AdS$_{D+1} \times X$ with the $U(1)$ an isometry on $X$.
Then, the corresponding symmetry operator is obtained from a solitonic configuration constructed from a diffeomorphism of $X$.
As a soliton, we again expect there to be a DBI-like action. That said, the field content on the brane deserves further study.

Consider next the case of $G$ a non-abelian Lie group with a single connected component (i.e., all elements can be continuously connected to the identity). In this case the symmetry operators will be labeled by elements of $G$. These naturally act on objects in $\text{Rep}(G)$, i.e., representations of $G$. The only change is that now, the symmetry current $j_{cd}$
will transform in a representation of the Lie algebra $\mathfrak{g} = \mathrm{Lie}(G)$, so the generator of interest will also be labeled by a
parameter $\mathfrak{t}^{cd} \in \mathfrak{g}$:
\begin{equation}
\mathcal{N}_{\mathfrak{t}} = \exp \left(2 \pi i \mathfrak{t}^{cd} \underset{Y_{D-1}}{\int} j_{cd} \right),
\end{equation}
in the obvious notation.

We can also attempt to mimic the steps taken in the $U(1)$ case to construct a bulk dual object.
The main difference is that now, our equation of motion involves a covariant derivative:
\begin{equation}
d_{A} F^{\text{dual}} = J_D,
\end{equation}
and as such, we cannot simply use Stokes' theorem to replace the integral over the boundary current by a $D$-chain which terminates on $Y_{D-1}$. So, on general grounds, we expect the ``naive''
$\mathcal{N}_{\mathfrak{t}}$ to also be dressed by $\mathcal{Z}$, as in line (\ref{Ntildez}). Said differently, the candidate topological operator in the bulk does not fully detach from the boundary.

This is in accord with the fact that if we do attempt to define an object purely in the bulk, then we should introduce a codimension $2$
vortex labeled by the conjugacy class of $\exp(2 \pi i \mathfrak{t})$, namely a Gukov-Witten operator
\cite{Gukov:2006jk}.\footnote{We thank H. Ooguri for a comment on this point.} One can obtain this by averaging the ``naive defect'' over gauge orbits, as in \cite{Cordova:2022rer}. It is also worth noting (see \cite{Gaiotto:2014kfa}) that in pure gauge theory, the only topological Gukov-Witten operators are associated with elements in the center of the Lie group $G$. For further discussion, especially in the context of gravity, see reference \cite{Heidenreich:2021xpr}.

Of course, on top of all of these complications we also expect there to be additional DBI-like contributions to the worldvolume action of $\widetilde{\mathcal{N}}$. For example, suppose that $G$ arises as a continuous non-abelian isometry group of some $X$ in a background of the form AdS$_{D+1} \times X$. Then, the brane worldvolume action will again descend from a solitonic configuration associated with the diffeomorphisms of $X$.

\section{Proofs with Splittability Revisited}\label{sec:SPLIT}

The line of argument just presented is somewhat different from that in
reference \cite{Harlow:2018jwu, Harlow:2018tng}, which concentrates on $0$-form symmetries,
as well as the dimensional reduction of $p$-form symmetries to $0$-form symmetries.
For ease of exposition we focus on the case of $0$-form symmetries,
since the other cases follow from similar steps to those presented in \cite{Harlow:2018jwu, Harlow:2018tng}.

Now, one of the central ideas there is to start with a spatial
region of the CFT $R$, and to consider a favorable
splitting into smaller regions $R_1,...,R_m$ such that the entanglement
wedge for any individual region $R_{i}$ is sufficiently close to the conformal
boundary of AdS. Doing so, one can then consider an operator deep in the
interior of AdS, and since no single region penetrates that far, a symmetry
operator confined to the region $R_{i}$ cannot properly act in the bulk. This
provides another way to argue for the absence of global symmetries in AdS. Of
course, a potential loophole in this argument is that one must somehow argue
that a topological operator of the boundary theory \textquotedblleft
cares\textquotedblright\ about the metric dependent structure of an
entanglement wedge in the first place. Our argument based on coarse graining
considerations shows why the gravity dual of operators such as $\mathcal{N}$
are still sensitive to the entanglement wedge.

Another subtlety in this line of reasoning is that in the case of a
non-invertible symmetry, we cannot simply take products of operators
to get another symmetry generator. Rather, we often get sums of symmetry operators, i.e.,
there is a non-trivial fusion rule.
In what follows we assume that the fusion rule only involves condensation defects,
i.e., objects of lower-dimensional support (see e.g., \cite{Gaiotto:2019xmp}).
The main reason to make this technical assumption is that there is still a
notion of a single ``big object'' which is present in the boundary theory and in the bulk.

Finally, we also face the issue of how to define the $\mathcal{N}[R_{j}]$ in the
first place; when $\partial R_{j}\neq0$, we generically expect the TFT on
$R_{j}$ to support edge modes. In principle these could either be gapless or
gapped degrees of freedom depending on the choice of boundary conditions for
the TFT for the symmetry operator.

The construction just provided in the previous section provides a
resolution for these issues. First of all, given a TFT\ on a region
$R$, we can introduce a \textquotedblleft dressed\textquotedblright\ operator
where we explicitly include the edge modes in question. We write this as:\footnote{For another perspective on splittability versus
concatenation of such topological operators, see reference \cite{Benedetti:2022zbb}.}
\begin{equation}
\overline{\mathcal{N}}[R]=\mathcal{N}[R]\mathcal{E}\lbrack\partial R],
\end{equation}
where $\mathcal{N}[R]$ refers to the path integral over the bulk TFT\ fields
on $R$, and $\mathcal{E}\lbrack\partial R]$ refers to the path integral over
any edge modes.\footnote{As an example, consider 3D\ Chern-Simons coupled to a
2D\ chiral CFT.}

\begin{figure}
\centering
\scalebox{0.8}{
\begin{tikzpicture}
	\begin{pgfonlayer}{nodelayer}
		\node [style=none] (0) at (-2.75, 2) {};
		\node [style=none] (1) at (-2.75, 0) {};
		\node [style=none] (2) at (-2.75, -2) {};
		\node [style=none] (3) at (-2.25, 2) {};
		\node [style=none] (4) at (-2.25, 0) {};
		\node [style=none] (5) at (-2.25, -2) {};
		\node [style=none] (6) at (-1.75, 2) {};
		\node [style=none] (7) at (-1.75, 0) {};
		\node [style=none] (8) at (-1.75, -2) {};
		\node [style=none] (9) at (-1.25, 2) {};
		\node [style=none] (10) at (-1.25, 0) {};
		\node [style=none] (11) at (-1.25, -2) {};
		\node [style=none] (12) at (-0.75, 2) {};
		\node [style=none] (13) at (-0.75, 0) {};
		\node [style=none] (14) at (-0.75, -2) {};
		\node [style=none] (15) at (-0.25, 2) {};
		\node [style=none] (16) at (-0.25, 0) {};
		\node [style=none] (17) at (-0.25, -2) {};
		\node [style=none] (18) at (0.25, 2) {};
		\node [style=none] (19) at (0.25, 0) {};
		\node [style=none] (20) at (0.25, -2) {};
		\node [style=none] (21) at (0.75, 2) {};
		\node [style=none] (22) at (0.75, 0) {};
		\node [style=none] (23) at (0.75, -2) {};
		\node [style=none] (24) at (1.25, 2) {};
		\node [style=none] (25) at (1.25, 0) {};
		\node [style=none] (26) at (1.25, -2) {};
		\node [style=none] (27) at (1.75, 2) {};
		\node [style=none] (28) at (1.75, 0) {};
		\node [style=none] (29) at (1.75, -2) {};
		\node [style=none] (30) at (-3.75, 2) {};
		\node [style=none] (31) at (-3.75, 0) {};
		\node [style=none] (32) at (-3.75, -2) {};
		\node [style=none] (33) at (-3.25, 2) {};
		\node [style=none] (34) at (-3.25, 0) {};
		\node [style=none] (35) at (-3.25, -2) {};
		\node [style=none] (36) at (3.25, 0) {};
		\node [style=none] (37) at (4.25, 0) {};
		\node [style=none] (38) at (11.25, 2) {};
		\node [style=none] (39) at (11.25, 0) {};
		\node [style=none] (40) at (11.25, -2) {};
		\node [style=none] (41) at (5.75, 2) {};
		\node [style=none] (42) at (5.75, 0) {};
		\node [style=none] (43) at (5.75, -2) {};
		\node [style=none] (44) at (8.5, 0) {3D Chern-Simons Theory};
		\node [style=none] (45) at (-1, -2.625) {Left / Right moving 2D CFTs};
		\node [style=none] (46) at (3.75, -0.5) {pairwise};
		\node [style=none] (47) at (3.75, 0.5) {gap};
		\node [style=none] (48) at (5.75, -2.625) {Chiral BC};
		\node [style=none] (49) at (11.25, -2.625) {Anti-Chiral BC};
		\node [style=none] (50) at (5.75, -3) {};
	\end{pgfonlayer}
	\begin{pgfonlayer}{edgelayer}
		\draw [style=ArrowLineRight] (0.center) to (1.center);
		\draw [style=ArrowLineRight] (5.center) to (4.center);
		\draw [style=ThickLine] (0.center) to (2.center);
		\draw [style=ThickLine] (5.center) to (3.center);
		\draw [style=ArrowLineRight] (6.center) to (7.center);
		\draw [style=ArrowLineRight] (11.center) to (10.center);
		\draw [style=ThickLine] (6.center) to (8.center);
		\draw [style=ThickLine] (11.center) to (9.center);
		\draw [style=ArrowLineRight] (12.center) to (13.center);
		\draw [style=ArrowLineRight] (17.center) to (16.center);
		\draw [style=ThickLine] (12.center) to (14.center);
		\draw [style=ThickLine] (17.center) to (15.center);
		\draw [style=ArrowLineRight] (18.center) to (19.center);
		\draw [style=ArrowLineRight] (23.center) to (22.center);
		\draw [style=ThickLine] (18.center) to (20.center);
		\draw [style=ThickLine] (23.center) to (21.center);
		\draw [style=ArrowLineRight] (24.center) to (25.center);
		\draw [style=ArrowLineRight] (29.center) to (28.center);
		\draw [style=ThickLine] (24.center) to (26.center);
		\draw [style=ThickLine] (29.center) to (27.center);
		\draw [style=ArrowLineRight] (30.center) to (31.center);
		\draw [style=ArrowLineRight] (35.center) to (34.center);
		\draw [style=ThickLine] (30.center) to (32.center);
		\draw [style=ThickLine] (35.center) to (33.center);
		\draw [style=ArrowLineRight] (36.center) to (37.center);
		\draw [style=ArrowLineRight] (40.center) to (39.center);
		\draw [style=ThickLine] (40.center) to (38.center);
		\draw [style=ArrowLineRight] (41.center) to (42.center);
		\draw [style=ThickLine] (41.center) to (43.center);
	\end{pgfonlayer}
\end{tikzpicture}}
\caption{LHS: System of 2D coupled wires. Arrows indicate left / right movers. RHS: Pairwise gapping out left and right movers results in a 3D Chern-Simons theory confined between the remaining wires which impose chiral / anti-chiral boundary conditions (BC). }
\label{fig:Wires}
\end{figure}
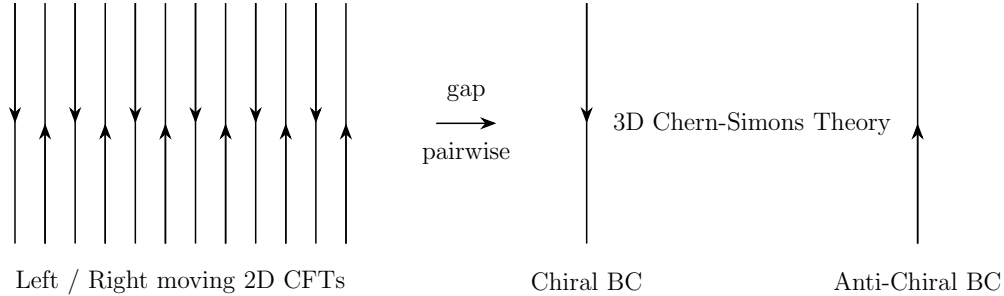

Suppose now that we have a region $R$ and we partition it up into a collection
of disjoint regions $\{R_{i}\}_{i}$ which nevertheless share common
boundaries. We would like to understand the relation between
$\overline{\mathcal{N}}[R]$ and $\overline{\mathcal{N}}[R_{1}%
]...\overline{\mathcal{N}}[R_{m}]$. Observe that in the product over the
split factors we have a collection of edge modes. The bulk perspective
indicates that we can fuse these edge modes together, and in so doing
integrate them out. At this point it is helpful to recall the coupled wires
construction of \cite{PhysRevLett.88.036401, PhysRevB.89.085101} which shows
how to explicitly carry this out for a collection of certain 2D\ CFTs where
neighboring left- and right-moving degrees of freedom are pairwise gapped out,
leaving behind a 3D bulk Chern-Simons theory with chiral / anti-chiral modes
on the very left and right of the system (see figure \ref{fig:Wires}).
We expect that something similar holds far more generally. This is indeed in accord with
holographic considerations where we view the $\widetilde{\mathcal{N}}$
operators as creating a brane in the bulk; edge modes for neighboring branes /
anti-branes condense, fusing the original configuration to a single large brane.

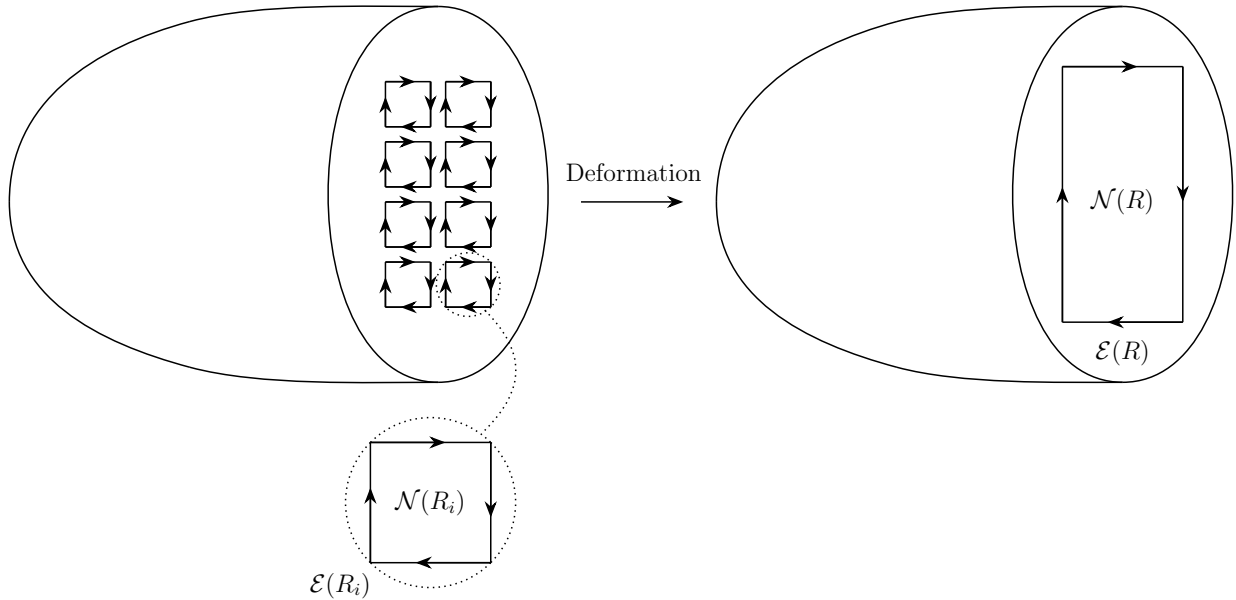
\begin{figure}
\centering
\scalebox{0.8}{
\begin{tikzpicture}
	\begin{pgfonlayer}{nodelayer}
		\node [style=none] (0) at (0.125, 3.25) {};
		\node [style=none] (1) at (0.125, -3) {};
		\node [style=none] (2) at (-4, 3) {};
		\node [style=none] (3) at (-4, -2.75) {};
		\node [style=none] (4) at (-7, 0) {};
		\node [style=none] (77) at (-1, -4) {};
		\node [style=none] (78) at (0.25, -4) {};
		\node [style=none] (79) at (1, -4) {};
		\node [style=none] (80) at (1, -5.25) {};
		\node [style=none] (81) at (1, -6) {};
		\node [style=none] (82) at (-0.25, -6) {};
		\node [style=none] (83) at (-1, -6) {};
		\node [style=none] (84) at (-1, -4.75) {};
		\node [style=none] (85) at (0, -5) {$\mathcal{N}(R_i)$};
		\node [style=none] (86) at (-1.5, -6.375) {$\mathcal{E}(R_i)$};
		\node [style=none] (87) at (2.5, 0) {};
		\node [style=none] (88) at (4.2, 0) {};
		\node [style=none] (89) at (11.5, 3.25) {};
		\node [style=none] (90) at (11.5, -3) {};
		\node [style=none] (91) at (7.75, 3) {};
		\node [style=none] (92) at (7.75, -2.75) {};
		\node [style=none] (93) at (4.75, 0) {};
		\node [style=none] (142) at (10.5, 2.25) {};
		\node [style=none] (143) at (12.5, 2.25) {};
		\node [style=none] (144) at (12.5, -2) {};
		\node [style=none] (145) at (10.5, -2) {};
		\node [style=none] (146) at (10.5, 0.25) {};
		\node [style=none] (147) at (12.5, 0) {};
		\node [style=none] (148) at (11.75, 2.25) {};
		\node [style=none] (149) at (11.25, -2) {};
		\node [style=none] (150) at (11.5, 0) {$\mathcal{N}(R)$};
		\node [style=none] (151) at (11.5, -2.5) {$\mathcal{E}(R)$};
		\node [style=none] (152) at (3.375, 0.5) {Deformation};
		\node [style=none] (214) at (0.25, -2) {};
		\node [style=none] (215) at (0, -2) {};
		\node [style=none] (216) at (0, -1.125) {};	
		\node [style=none] (217) at (0.25, -1.75) {};
		\node [style=none] (218) at (1, -1.75) {};
		\node [style=none] (219) at (1, -1) {};
		\node [style=none] (220) at (0.25, -1) {};
		\node [style=none] (221) at (-0.25, -1.5) {};
		\node [style=none] (222) at (0.9, -1.8) {};
		\node [style=none] (223) at (0.9, -3.8) {};
		\node [style=none] (224) at (-1.25, -4.75) {};
		\node [style=none] (225) at (3, -7) {};
		\node [style=none] (226) at (-0.75, 2) {};
		\node [style=none] (227) at (-0.25, 2) {};
		\node [style=none] (228) at (0, 2) {};
		\node [style=none] (229) at (0, 1.5) {};
		\node [style=none] (230) at (0, 1.25) {};
		\node [style=none] (231) at (-0.5, 1.25) {};
		\node [style=none] (232) at (-0.75, 1.25) {};
		\node [style=none] (233) at (-0.75, 1.75) {};
		\node [style=none] (234) at (-0.75, 1.25) {};
		\node [style=none] (236) at (0, 1.25) {};
		\node [style=none] (238) at (0, 2) {};
		\node [style=none] (239) at (0, 1.25) {};
		\node [style=none] (241) at (0, 1.25) {};
		\node [style=none] (242) at (0.25, 2) {};
		\node [style=none] (243) at (0.75, 2) {};
		\node [style=none] (244) at (1, 2) {};
		\node [style=none] (245) at (1, 1.5) {};
		\node [style=none] (246) at (1, 1.25) {};
		\node [style=none] (247) at (0.5, 1.25) {};
		\node [style=none] (248) at (0.25, 1.25) {};
		\node [style=none] (249) at (0.25, 1.75) {};
		\node [style=none] (250) at (0.25, 1.25) {};
		\node [style=none] (251) at (1, 1.25) {};
		\node [style=none] (252) at (1, 2) {};
		\node [style=none] (253) at (1, 1.25) {};
		\node [style=none] (254) at (1, 1.25) {};
		\node [style=none] (255) at (-0.75, 1) {};
		\node [style=none] (256) at (-0.25, 1) {};
		\node [style=none] (257) at (0, 1) {};
		\node [style=none] (258) at (0, 0.5) {};
		\node [style=none] (259) at (0, 0.25) {};
		\node [style=none] (260) at (-0.5, 0.25) {};
		\node [style=none] (261) at (-0.75, 0.25) {};
		\node [style=none] (262) at (-0.75, 0.75) {};
		\node [style=none] (263) at (-0.75, 0.25) {};
		\node [style=none] (264) at (0, 0.25) {};
		\node [style=none] (265) at (0, 1) {};
		\node [style=none] (266) at (0, 0.25) {};
		\node [style=none] (267) at (0, 0.25) {};
		\node [style=none] (268) at (0.25, 1) {};
		\node [style=none] (269) at (0.75, 1) {};
		\node [style=none] (270) at (1, 1) {};
		\node [style=none] (271) at (1, 0.5) {};
		\node [style=none] (272) at (1, 0.25) {};
		\node [style=none] (273) at (0.5, 0.25) {};
		\node [style=none] (274) at (0.25, 0.25) {};
		\node [style=none] (275) at (0.25, 0.75) {};
		\node [style=none] (276) at (0.25, 0.25) {};
		\node [style=none] (277) at (1, 0.25) {};
		\node [style=none] (278) at (1, 1) {};
		\node [style=none] (279) at (1, 0.25) {};
		\node [style=none] (280) at (1, 0.25) {};
		\node [style=none] (281) at (-0.75, 0) {};
		\node [style=none] (282) at (-0.25, 0) {};
		\node [style=none] (283) at (0, 0) {};
		\node [style=none] (284) at (0, -0.5) {};
		\node [style=none] (285) at (0, -0.75) {};
		\node [style=none] (286) at (-0.5, -0.75) {};
		\node [style=none] (287) at (-0.75, -0.75) {};
		\node [style=none] (288) at (-0.75, -0.25) {};
		\node [style=none] (289) at (-0.75, -0.75) {};
		\node [style=none] (290) at (0, -0.75) {};
		\node [style=none] (291) at (0, 0) {};
		\node [style=none] (292) at (0, -0.75) {};
		\node [style=none] (293) at (0, -0.75) {};
		\node [style=none] (294) at (0.25, 0) {};
		\node [style=none] (295) at (0.75, 0) {};
		\node [style=none] (296) at (1, 0) {};
		\node [style=none] (297) at (1, -0.5) {};
		\node [style=none] (298) at (1, -0.75) {};
		\node [style=none] (299) at (0.5, -0.75) {};
		\node [style=none] (300) at (0.25, -0.75) {};
		\node [style=none] (301) at (0.25, -0.25) {};
		\node [style=none] (302) at (0.25, -0.75) {};
		\node [style=none] (303) at (1, -0.75) {};
		\node [style=none] (304) at (1, 0) {};
		\node [style=none] (305) at (1, -0.75) {};
		\node [style=none] (306) at (1, -0.75) {};
		\node [style=none] (307) at (-0.75, -1) {};
		\node [style=none] (308) at (-0.25, -1) {};
		\node [style=none] (309) at (0, -1) {};
		\node [style=none] (310) at (0, -1.5) {};
		\node [style=none] (311) at (0, -1.75) {};
		\node [style=none] (312) at (-0.5, -1.75) {};
		\node [style=none] (313) at (-0.75, -1.75) {};
		\node [style=none] (314) at (-0.75, -1.25) {};
		\node [style=none] (315) at (-0.75, -1.75) {};
		\node [style=none] (316) at (0, -1.75) {};
		\node [style=none] (317) at (0, -1) {};
		\node [style=none] (318) at (0, -1.75) {};
		\node [style=none] (319) at (0, -1.75) {};
		\node [style=none] (320) at (0.25, -1) {};
		\node [style=none] (321) at (0.75, -1) {};
		\node [style=none] (322) at (1, -1) {};
		\node [style=none] (323) at (1, -1.5) {};
		\node [style=none] (324) at (1, -1.75) {};
		\node [style=none] (325) at (0.5, -1.75) {};
		\node [style=none] (326) at (0.25, -1.75) {};
		\node [style=none] (327) at (0.25, -1.25) {};
		\node [style=none] (328) at (0.25, -1.75) {};
		\node [style=none] (329) at (1, -1.75) {};
		\node [style=none] (330) at (1, -1) {};
		\node [style=none] (331) at (1, -1.75) {};
		\node [style=none] (332) at (1, -1.75) {};
	\end{pgfonlayer}
	\begin{pgfonlayer}{edgelayer}
		\draw [style=ThickLine, in=-15, out=180, looseness=0.75] (1.center) to (3.center);
		\draw [style=ThickLine, in=-90, out=165] (3.center) to (4.center);
		\draw [style=ThickLine, in=-165, out=90] (4.center) to (2.center);
		\draw [style=ThickLine, in=-180, out=15, looseness=0.75] (2.center) to (0.center);
		\draw [style=ThickLine, bend right=90] (0.center) to (1.center);
		\draw [style=ThickLine, bend left=90] (0.center) to (1.center);
		\draw [style=ArrowLineRight] (77.center) to (78.center);
		\draw [style=ThickLine] (77.center) to (79.center);
		\draw [style=ArrowLineRight] (79.center) to (80.center);
		\draw [style=ArrowLineRight] (81.center) to (82.center);
		\draw [style=ArrowLineRight] (83.center) to (84.center);
		\draw [style=ThickLine] (79.center) to (81.center);
		\draw [style=ThickLine] (81.center) to (83.center);
		\draw [style=ThickLine] (83.center) to (77.center);
		\draw [style=ArrowLineRight] (87.center) to (88.center);
		\draw [style=ThickLine, in=-15, out=180, looseness=0.75] (90.center) to (92.center);
		\draw [style=ThickLine, in=-90, out=165] (92.center) to (93.center);
		\draw [style=ThickLine, in=-165, out=90] (93.center) to (91.center);
		\draw [style=ThickLine, in=-180, out=15, looseness=0.75] (91.center) to (89.center);
		\draw [style=ThickLine, bend right=90] (89.center) to (90.center);
		\draw [style=ThickLine, bend left=90] (89.center) to (90.center);
		\draw [style=ArrowLineRight] (145.center) to (146.center);
		\draw [style=ArrowLineRight] (143.center) to (147.center);
		\draw [style=ArrowLineRight] (142.center) to (148.center);
		\draw [style=ArrowLineRight] (144.center) to (149.center);
		\draw [style=ThickLine] (145.center) to (142.center);
		\draw [style=ThickLine] (142.center) to (143.center);
		\draw [style=ThickLine] (143.center) to (144.center);
		\draw [style=ThickLine] (144.center) to (145.center);
		\draw [style=DottedLine, bend left=45] (220.center) to (219.center);
		\draw [style=DottedLine, bend left=45] (219.center) to (218.center);
		\draw [style=DottedLine, bend left=45] (218.center) to (217.center);
		\draw [style=DottedLine, bend left=45] (217.center) to (220.center);
		\draw [style=DottedLine, bend left=45] (77.center) to (79.center);
		\draw [style=DottedLine, bend left=45] (79.center) to (81.center);
		\draw [style=DottedLine, bend left=45] (81.center) to (83.center);
		\draw [style=DottedLine, bend right=45] (77.center) to (83.center);
		\draw [style=DottedLine, bend left=45, looseness=1.25] (222.center) to (223.center);
		\draw [style=ArrowLineRight] (226.center) to (227.center);
		\draw [style=ThickLine] (226.center) to (228.center);
		\draw [style=ArrowLineRight] (228.center) to (229.center);
		\draw [style=ArrowLineRight] (230.center) to (231.center);
		\draw [style=ArrowLineRight] (232.center) to (233.center);
		\draw [style=ThickLine] (228.center) to (230.center);
		\draw [style=ThickLine] (230.center) to (232.center);
		\draw [style=ThickLine] (232.center) to (226.center);
		\draw [style=ThickLine] (234.center) to (236.center);
		\draw [style=ThickLine] (239.center) to (238.center);
		\draw [style=ArrowLineRight] (242.center) to (243.center);
		\draw [style=ThickLine] (242.center) to (244.center);
		\draw [style=ArrowLineRight] (244.center) to (245.center);
		\draw [style=ArrowLineRight] (246.center) to (247.center);
		\draw [style=ArrowLineRight] (248.center) to (249.center);
		\draw [style=ThickLine] (244.center) to (246.center);
		\draw [style=ThickLine] (246.center) to (248.center);
		\draw [style=ThickLine] (248.center) to (242.center);
		\draw [style=ThickLine] (250.center) to (251.center);
		\draw [style=ThickLine] (253.center) to (252.center);
		\draw [style=ArrowLineRight] (255.center) to (256.center);
		\draw [style=ThickLine] (255.center) to (257.center);
		\draw [style=ArrowLineRight] (257.center) to (258.center);
		\draw [style=ArrowLineRight] (259.center) to (260.center);
		\draw [style=ArrowLineRight] (261.center) to (262.center);
		\draw [style=ThickLine] (257.center) to (259.center);
		\draw [style=ThickLine] (259.center) to (261.center);
		\draw [style=ThickLine] (261.center) to (255.center);
		\draw [style=ThickLine] (263.center) to (264.center);
		\draw [style=ThickLine] (266.center) to (265.center);
		\draw [style=ArrowLineRight] (268.center) to (269.center);
		\draw [style=ThickLine] (268.center) to (270.center);
		\draw [style=ArrowLineRight] (270.center) to (271.center);
		\draw [style=ArrowLineRight] (272.center) to (273.center);
		\draw [style=ArrowLineRight] (274.center) to (275.center);
		\draw [style=ThickLine] (270.center) to (272.center);
		\draw [style=ThickLine] (272.center) to (274.center);
		\draw [style=ThickLine] (274.center) to (268.center);
		\draw [style=ThickLine] (276.center) to (277.center);
		\draw [style=ThickLine] (279.center) to (278.center);
		\draw [style=ArrowLineRight] (281.center) to (282.center);
		\draw [style=ThickLine] (281.center) to (283.center);
		\draw [style=ArrowLineRight] (283.center) to (284.center);
		\draw [style=ArrowLineRight] (285.center) to (286.center);
		\draw [style=ArrowLineRight] (287.center) to (288.center);
		\draw [style=ThickLine] (283.center) to (285.center);
		\draw [style=ThickLine] (285.center) to (287.center);
		\draw [style=ThickLine] (287.center) to (281.center);
		\draw [style=ThickLine] (289.center) to (290.center);
		\draw [style=ThickLine] (292.center) to (291.center);
		\draw [style=ArrowLineRight] (294.center) to (295.center);
		\draw [style=ThickLine] (294.center) to (296.center);
		\draw [style=ArrowLineRight] (296.center) to (297.center);
		\draw [style=ArrowLineRight] (298.center) to (299.center);
		\draw [style=ArrowLineRight] (300.center) to (301.center);
		\draw [style=ThickLine] (296.center) to (298.center);
		\draw [style=ThickLine] (298.center) to (300.center);
		\draw [style=ThickLine] (300.center) to (294.center);
		\draw [style=ThickLine] (302.center) to (303.center);
		\draw [style=ThickLine] (305.center) to (304.center);
		\draw [style=ArrowLineRight] (307.center) to (308.center);
		\draw [style=ThickLine] (307.center) to (309.center);
		\draw [style=ArrowLineRight] (309.center) to (310.center);
		\draw [style=ArrowLineRight] (311.center) to (312.center);
		\draw [style=ArrowLineRight] (313.center) to (314.center);
		\draw [style=ThickLine] (309.center) to (311.center);
		\draw [style=ThickLine] (311.center) to (313.center);
		\draw [style=ThickLine] (313.center) to (307.center);
		\draw [style=ThickLine] (315.center) to (316.center);
		\draw [style=ThickLine] (318.center) to (317.center);
		\draw [style=ArrowLineRight] (320.center) to (321.center);
		\draw [style=ThickLine] (320.center) to (322.center);
		\draw [style=ArrowLineRight] (322.center) to (323.center);
		\draw [style=ArrowLineRight] (324.center) to (325.center);
		\draw [style=ArrowLineRight] (326.center) to (327.center);
		\draw [style=ThickLine] (322.center) to (324.center);
		\draw [style=ThickLine] (324.center) to (326.center);
		\draw [style=ThickLine] (326.center) to (320.center);
		\draw [style=ThickLine] (328.center) to (329.center);
		\draw [style=ThickLine] (331.center) to (330.center);
	\end{pgfonlayer}
\end{tikzpicture}}
\caption{On the left we depict in the CFT$_D$ a collection of the $\mathcal{N}(R_i)$ supported on compact subspaces $R_i$ with boundary $\partial R_i$ which are dressed by non-topological edgemodes $\mathcal{E}(\partial R_i)$. Whenever these constitute a tiling we can turn on deformations to gap out edgemodes pairwise, resulting in a larger symmetry operator, with possibly some edgemodes remaining.}
\label{fig:Stokes}
\end{figure}

With this, we have a deformation of the original TFT\ / edge mode system
to the single dressed operator:
\begin{equation}
\overline{\mathcal{N}}[R_{1}]...\overline{\mathcal{N}}[R_{m}] \rightsquigarrow \overline{\mathcal{N}}[R].
\end{equation}
See figure \ref{fig:Stokes} for a depiction of this merging procedure.
One can then proceed with the same style of proof as in \cite{Harlow:2018jwu,
Harlow:2018tng}, now extended to the case of non-invertible symmetries.

\section{Further Comments}

In this section we provide some further, more speculative comments which naturally extend the considerations just presented. We begin by extending the notion of a higher-form symmetry to the case of lower-form symmetry, i.e., a $(-m)$-form global symmetry for $m \geq 2$, and also discuss its holographic interpretation. Next, we discuss how our considerations extend to more general spacetimes with
holographic screens.

\subsection{Lower-Form Symmetries}

It is of course tempting to extend our discussion to cover $(-m)$-form
symmetries. One reason for doing so is to ask whether the bulk AdS space can
support bulk parameters, i.e., can we have a global $(-1)$-form symmetry in
the bulk? Formally speaking this would be a \textquotedblleft$(-2)$-form
symmetry\textquotedblright\ in the boundary CFT. This is of direct interest in
the context of a number of questions in quantum gravity, i.e., can quantum
gravity support bulk parameters at all? See e.g., \cite{Marolf:2020xie,
McNamara:2020uza, Debray:2021vob, Heckman:2021vzx, Baume:2023kkf} for some different
perspectives. The construction of such objects in the dual CFT$_{D}$
sounds puzzling since it seems to require a topological operator which fills
more than the spacetime dimensions of the system.

We now propose to give a formal definition of a $(-m)$-form symmetry for any
QFT, which we can of course apply to the special case of holographic CFTs.
As a general comment, we have deferred discussion of this
case because it is necessarily somewhat more speculative.

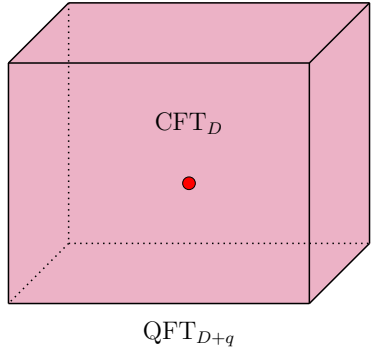
\begin{figure}
\centering
\scalebox{0.8}{
\begin{tikzpicture}
	\begin{pgfonlayer}{nodelayer}
		\node [style=none] (0) at (-3, 1) {};
		\node [style=none] (1) at (-2, 2) {};
		\node [style=none] (2) at (2, 1) {};
		\node [style=none] (3) at (3, 2) {};
		\node [style=none] (4) at (-3, -3) {};
		\node [style=none] (5) at (-2, -2) {};
		\node [style=none] (6) at (2, -3) {};
		\node [style=none] (7) at (3, -2) {};
		\node [style=none] (10) at (0, -2.5) {};
		\node [style=none] (12) at (0, 0) {CFT$_D$};
		\node [style=none] (13) at (0, -3.5) {QFT$_{D+q}$};
		\node [style=none] (14) at (0, 1) {};
		\node [style=none] (15) at (0, 1.5) {};
		\node [style=CircleRed] (16) at (0, -1) {};
		\node [style=none] (17) at (0, -1.5) {};
	\end{pgfonlayer}
	\begin{pgfonlayer}{edgelayer}
		\filldraw[fill=purple!30, draw=purple!30]  (-3, 1) -- (-2, 2) -- (3, 2) -- (3, 1) -- cycle;
        \filldraw[fill=purple!30, draw=purple!30]  (-3, -3) -- (-3, -2) -- (3, -2) -- (2, -3) -- cycle;
        \filldraw[fill=purple!30, draw=purple!30]  (-3, -2) -- (-3, 1) -- (3, 1) -- (3, -2) -- cycle;
		\draw [style=ThickLine] (0.center) to (4.center);
		\draw [style=ThickLine] (2.center) to (6.center);
		\draw [style=ThickLine] (3.center) to (7.center);
		\draw [style=ThickLine] (6.center) to (4.center);
		\draw [style=ThickLine] (0.center) to (2.center);
		\draw [style=ThickLine] (2.center) to (3.center);
		\draw [style=ThickLine] (3.center) to (1.center);
		\draw [style=ThickLine] (1.center) to (0.center);
		\draw [style=ThickLine] (6.center) to (7.center);
		\draw [style=DottedLine] (5.center) to (7.center);
		\draw [style=DottedLine] (5.center) to (4.center);
		\draw [style=DottedLine] (1.center) to (5.center);
	\end{pgfonlayer}
\end{tikzpicture}
}
\caption{We can construct $(-m)$-form symmetries for a CFT$_D$ by viewing it as a defect (red dot) inside of an ambient QFT$_{D+q}$ with
$q = m - 1$. In the higher dimensional setting, an ambient $(-1)$-form symmetry associated with a parameter of QFT$_{D+q}$ formally defines a $(-m)$-form symmetry of the defect system. If the CFT$_D$ has a gravity dual, this yields a bulk $(-m+1)$-form symmetry.}
\label{fig:BCFT}
\end{figure}

We propose to view the $D$-dimensional QFT$_{D}$ as a defect in a higher-dimensional QFT$_{D+q}$ (see figure \ref{fig:BCFT}).
Setting $q \equiv m-1$, observe that a spacetime filling $(-1)$-form symmetry operator of QFT$_{D+q}$ can be interpreted, in the QFT$_{D}$ theory, as a $(-m)$-form symmetry.\footnote{Why not simply demand that we have a $p$-form symmetry operator
of the bulk which fills all the worldvolume directions of the QFT$_D$? If we can move the
topological operator off of the QFT$_D$ it is unclear whether the QFT$_D$ will continue to see the
symmetry operator in the first place. Indeed, otherwise this would likely lead to contradictions with the standard treatment of $(-1)$-form symmetries.}
It is also natural to consider the entwinement and nested categorical
structure of these symmetries. A related comment is that this sort of symmetry
inheritance arises quite naturally in many bulk / boundary systems, including
some stringy constructions (see e.g., the recent discussion in reference \cite{Acharya:2023bth}).
It would also be natural to ask whether lower-form symmetries can be defined intrinsically or must implicitly always make reference to a bulk QFT.

Before proceeding to the holographic interpretation, let us mention that at least in certain circumstances one might wish to entertain a different notion of lower-form symmetries based on taking a field profile which is sensitive to topological structures in the target space, i.e., topologically non-trivial field excursions in the QFT$_D$.\footnote{We thank J. McNamara and M. Montero for helpful comments on this point.}
At least in explicit stringy constructions of QFTs there is certainly overlap with our proposal
since we can consider ``moving the brane around'' in the ambient bulk QFT$_{D+q}$, but it would clearly be interesting to explore this generalization as well.

\begin{figure}
\centering
\scalebox{0.8}{
\begin{tikzpicture}
	\begin{pgfonlayer}{nodelayer}
		\node [style=none] (0) at (-3, 3) {};
		\node [style=none] (1) at (3, 3) {};
		\node [style=none] (2) at (2, 0) {};
		\node [style=none] (3) at (3, -3) {};
		\node [style=none] (4) at (-3, -3) {};
		\node [style=none] (5) at (-2, 0) {};
		\node [style=none] (6) at (-1.75, 0) {};
		\node [style=none] (7) at (-1.2, 0.6) {};
		\node [style=none] (8) at (1.2, 0.6) {};
		\node [style=none] (9) at (1.75, 0) {};
		\node [style=none] (10) at (-10, 3) {};
		\node [style=none] (11) at (-6, 3) {};
		\node [style=none] (12) at (-6, -3) {};
		\node [style=none] (13) at (-10, -3) {};
		\node [style=none] (14) at (-5.25, 0) {};
		\node [style=none] (15) at (-3.25, 0) {};
		\node [style=none] (16) at (-4.75, 0) {};
		\node [style=none] (17) at (-3.75, 0) {};
		\node [style=none] (18) at (-9.75, 0) {};
		\node [style=none] (19) at (-9.25, 0) {};
		\node [style=none] (20) at (-6.75, 0) {};
		\node [style=none] (21) at (-6.25, 0) {};
		\node [style=none] (22) at (-9, 3) {};
		\node [style=none] (23) at (-7, 3) {};
		\node [style=none] (24) at (-9, -3) {};
		\node [style=none] (25) at (-7, -3) {};
		\node [style=none] (26) at (-1, 0.75) {};
		\node [style=none] (27) at (1, 0.75) {};
		\node [style=none] (28) at (1.5, 3) {};
		\node [style=none] (29) at (-1.5, 3) {};
		\node [style=none] (30) at (-1.5, -3) {};
		\node [style=none] (31) at (1.5, -3) {};
		\node [style=none] (32) at (-8, -4.125) {\large (i)};
		\node [style=none] (33) at (0, -4.125) {\large (ii)};
		\node [style=none] (34) at (-4, -4.75) {};
		\node [style=none] (35) at (-11, 0) {$t=0$};
		\node [style=none] (36) at (2.75, 0) {$\tau=0$};
		\node [style=none] (39) at (-11.5, -3) {};
	\end{pgfonlayer}
	\begin{pgfonlayer}{edgelayer}
		\filldraw[fill=blue!30, draw=blue!30] (0,0) ellipse (2.0cm and 0.3cm);
		\filldraw[fill=blue!30, draw=blue!30] (-8,0) ellipse (2.0cm and 0.3cm);
		\filldraw[fill=blue!30, draw=blue!30] (-8,1.5) ellipse (2.0cm and 0.3cm);
		\filldraw[fill=blue!30, draw=blue!30] (-8,-1.5) ellipse (2.0cm and 0.3cm);
		\filldraw[fill=red!70, draw=red!70] (0,0.75) ellipse (1cm and 0.2cm);
		\filldraw[fill=red!70, draw=red!70] (-8,0) ellipse (1cm and 0.2cm);
		\filldraw[fill=blue!30, draw=blue!30] (0,-1.5) ellipse (2.35cm and 0.3cm);
		\filldraw[fill=blue!30, draw=blue!30] (0,1.5) ellipse (2.35cm and 0.3cm);
		\draw [style=ThickLine, in=90, out=-75] (0.center) to (5.center);
		\draw [style=ThickLine, in=75, out=-90] (5.center) to (4.center);
		\draw [style=ThickLine, in=-90, out=105] (3.center) to (2.center);
		\draw [style=ThickLine, in=-105, out=90] (2.center) to (1.center);
		\draw [style=ThickLine, bend left=90, looseness=0.25] (1.center) to (0.center);
		\draw [style=ThickLine, bend right=90, looseness=0.25] (4.center) to (3.center);
		\draw [style=ThickLine, bend left=90, looseness=0.25] (0.center) to (1.center);
		\draw [style=DottedLine, bend left=90, looseness=0.25] (4.center) to (3.center);
		\draw [style=ArrowLineRight] (6.center) to (7.center);
		\draw [style=ArrowLineRight] (9.center) to (8.center);
		\draw [style=ThickLine, bend left=90, looseness=0.25] (11.center) to (10.center);
		\draw [style=ThickLine, bend right=90, looseness=0.25] (13.center) to (12.center);
		\draw [style=ThickLine, bend left=90, looseness=0.25] (10.center) to (11.center);
		\draw [style=DottedLine, bend left=90, looseness=0.25] (13.center) to (12.center);
		\draw [style=ThickLine] (11.center) to (12.center);
		\draw [style=ThickLine] (10.center) to (13.center);
		\draw [style=ArrowLineRight] (18.center) to (19.center);
		\draw [style=ArrowLineRight] (21.center) to (20.center);
		\draw [style=DashedLine] (22.center) to (24.center);
		\draw [style=DashedLine] (23.center) to (25.center);
		\draw [style=DashedLine, in=90, out=-90] (29.center) to (26.center);
		\draw [style=DashedLine, in=90, out=-90] (26.center) to (30.center);
		\draw [style=DashedLine, in=90, out=-90] (27.center) to (31.center);
		\draw [style=DashedLine, in=90, out=-90] (28.center) to (27.center);
		\draw [style=DashedLine, bend left=270, looseness=0.25] (28.center) to (29.center);
		\draw [style=DashedLine, bend right=90, looseness=0.25] (29.center) to (28.center);
		\draw [style=DashedLine, bend left=90, looseness=0.25] (22.center) to (23.center);
		\draw [style=DashedLine, bend right=90, looseness=0.25] (24.center) to (25.center);
		\draw [style=DashedLine, bend right=90, looseness=0.25] (30.center) to (31.center);
		\draw [style=DashedLine, bend right=90, looseness=0.25] (22.center) to (23.center);
		\draw [style=DashedLine, bend left=90, looseness=0.25] (24.center) to (25.center);
		\draw [style=DashedLine, bend left=90, looseness=0.25] (30.center) to (31.center);
	\end{pgfonlayer}
\end{tikzpicture}
}
\caption{Depiction of bulk reconstruction in AdS/CFT and more general holographic spacetimes.
(i): AdS/CFT setup with leaves of constant time $t$ (blue). RG / bulk flow evolves cylindrical shells radially inwards. For example, leaves shrink in size (red). (ii): More general holographic spacetimes. In this case, the boundary theory is supported on a holographic screen.}
\label{fig:MoreGeneralSpacetimes}
\end{figure}
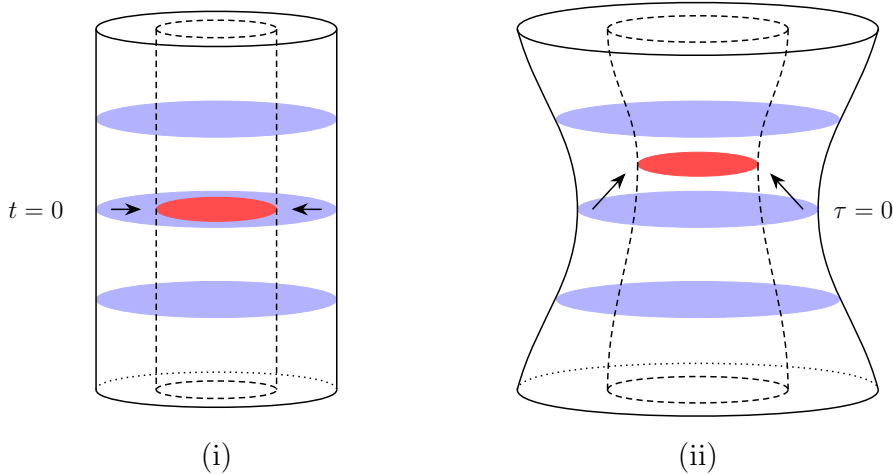

What then is the gravity dual of a $(-m)$-form symmetry operator? The most
conservative answer is to simply consider a CFT in dimension $D+q=D+m-1$ with
a semi-classical gravity dual on AdS$_{D+m}$. Then, the $(-m)$-form symmetry
of the defect QFT\ directly lifts to a codimension one wall of the ambient gravity
theory, much as in our treatment of $(-1)$-form symmetries.

On the other hand, if we start with a known AdS$_{D+1}$/CFT$_{D}$ pair and ask
about the fate of a bulk $(-1)$-form symmetry, we seem to require that our
boundary CFT\ can also be viewed as a defect in a QFT$_{D+q}$ decoupled from
gravity. In that setting, we need not (and should not!) demand that the
QFT$_{D+q}$ has its own semi-classical AdS dual. Rather,
we might have a spacetime of the form AdS$_{D+1} \times X$ with the
QFT$_{D+q}$ filling the AdS$_{D+1}$ factor and possibly some subspace of $X$.
It is also tempting to speculate that here, the appropriate notion of ``smearing'' involves coarse graining in the directions of the $X$ factor of AdS$_{D+1} \times X$. We leave a full treatment of these intriguing possibilities for future work.

\subsection{More General Spacetimes}

Similar considerations hold for a general spacetime that
admits a bulk-boundary duality with a suitable notion of entanglement wedge reconstruction and coarse-graining.
The surface-state correspondence provides this framework for any convex region
$M$ in a static spacetime wherein the dual boundary theory lives on $\partial M$  \cite{Miyaji:2015fia, Miyaji:2015yva},
see figure \ref{fig:MoreGeneralSpacetimes} for a depiction.\footnote{See \cite{Nomura:2016ikr,Nomura:2017fyh,Nomura:2017npr} for covariant generalizations wherein the boundary theory lives on the holographic screen \cite{Bousso:1999cb}. See also \cite{Strominger:2001pn, Alishahiha:2004md, Dong:2011uf} for other proposals for holography in general spacetimes.}
Coarse-graining is given by the bulk flow procedure \cite{Nomura:2018kji, Murdia:2020iac}. Note that there is also a notion of approximate locality in this boundary theory because the entanglement wedges become smaller closer to the boundary theory.
We can now repeat our argument establishing that a reconstructed $\widetilde{\mathcal{N}}$ sources some stress energy.
So, establishing the existence of entanglement wedge reconstruction in
more general spacetimes would also exclude global symmetries.


\section*{Acknowledgements}

We thank V. Balasubramanian, F. Baume, J. Calder\'{o}n-Infante, C.L. Kane, J. Kulp, J. McNamara,
M. Montero, E. Torres, A.P. Turner, C. Vafa, and X. Yu for helpful
discussions. We thank C. Akers, F. Apruzzi, J. McNamara, M. Montero,
Y. Nomura, H. Ooguri, and T. Rudelius for helpful comments on an earlier draft.
The work of JJH and CM is supported by DOE (HEP) Award
DE-SC0013528. The work of JJH and MH is supported in part by a University
Research Foundation grant at the University of Pennsylvania. The work of JJH
and MH is supported in part by BSF grant 2022100. The work of MH is also
supported by the Simons Foundation Collaboration grant \#724069 on
\textquotedblleft Special Holonomy in Geometry, Analysis and
Physics\textquotedblright.


\bibliographystyle{utphys}
\bibliography{NoGlobo}

\end{document}